\documentclass[aps,pre,letterpaper,superscriptaddress,endfloats,showpacs]{revtex4}

\usepackage{graphicx}
\usepackage{amssymb}

\def\lbullet{\mbox{\large $\bullet$}}
\def\ssquare{\mbox{\footnotesize $\square$}}
\def\sblacksquare{\mbox{\footnotesize $\blacksquare$}}

\def\lcirc{\mbox{\large $\circ$}}

\def\bfr{{\bf r}} 
\def\bfk{{\bf k}} 
\def\bfu{{\bf u}} 
\def\bfv{{\bf v}}

\def\bomega{\mbox{\boldmath $\omega$}}           
 
\def\bxi{\mbox{\boldmath $\xi$}} 
\def\bsxi{\mbox{\boldmath \tiny $\xi$}} 
\def\zhat{\hat{\bf z}}

\newcommand{\q}[1]{\fontencoding{U}\fontfamily
{bbm}\fontseries{m}\fontshape{n}\selectfont#1 \normalfont\normalsize}

\def\qa{{\textnormal{\q{A}}{\hskip -0.13cm}}} 
\def\qv{{\textnormal{\q{V}}{\hskip -0.13cm}}} 
\def\qvr{{\textnormal{\q{V}}}^{{\hskip -0.12cm} r}}
\def\qu{{\textnormal{\q{U}}{\hskip -0.13cm}}} 
\def\qusq{{\textnormal{\q{U}}}^{{\hskip -0.13cm}2}}
\def\qucube{{\textnormal{\q{U}}}^{{\hskip -0.13cm}3}}
\def\qac{{\textnormal{\q{A}}}^{{\hskip -0.2cm}*}}
\def\qap{{\textnormal{\q{A}}}^{{\hskip -0.2cm}\prime}}
\def\qapc{{\textnormal{\q{A}}}^{{\hskip -0.2cm}\prime*}}
\def\qvzero{{\textnormal{\q{V}}}{{\hskip -0.15cm}(0)}}
\def\qvtau{{\textnormal{\q{V}}}{{\hskip -0.15cm}(\tau)}}
\def\qvtwotau{{\textnormal{\q{V}}}{{\hskip -0.15cm}(2\tau)}}
\def\qvcubetau{{\textnormal{\q{V}}}^{{\hskip -0.13cm}3} (\tau)}
\def\qvcubetwotau{{\textnormal{\q{V}}}^{{\hskip -0.13cm}3} (2\tau)}
\def\qvcube{{\textnormal{\q{V}}}^{{\hskip -0.13cm}3}}

\def\qP{{\textnormal{\q{P}}{\hskip -0.13cm}}}
\def\qQ{{\textnormal{\q{Q}}{\hskip -0.13cm}}}     
\def\qR{{\textnormal{\q{R}}{\hskip -0.13cm}}}
\def\qS{{\textnormal{\q{S}}{\hskip -0.13cm}}}     

\begin{document}   
\title{Transport Coefficients for Stochastic Rotation Dynamics in Three 
Dimensions}
\author{E. T{\" uzel}}
\affiliation{School of Physics and Astronomy, 116 Church Street SE, 
University of Minnesota, Minneapolis, MN 55455, USA }
\affiliation{Supercomputing Institute, University of Minnesota, \\
599 Walter Library, 117 Pleasant Street S.E., \\
Minneapolis, MN 55455, USA}
\author{M. Strauss}
\affiliation{ Institut f\"ur Computeranwendungen 1, Universit\"at 
Stuttgart \\ Pfaffenwaldring 27, 70569 Stuttgart, Germany}
\author{T. Ihle}
\affiliation{ Institut f\"ur Computeranwendungen 1, Universit\"at 
Stuttgart \\ Pfaffenwaldring 27, 70569 Stuttgart, Germany}
\author{D.M. Kroll}
\affiliation{Supercomputing Institute, University of Minnesota, \\
599 Walter Library, 117 Pleasant Street S.E., \\
Minneapolis, MN 55455, USA}

\begin{abstract}

Explicit expressions for the transport coefficients of a recently introduced 
stochastic model for simulating fluctuating fluid dynamics are derived in 
three dimensions by means of Green-Kubo relations and simple kinetic arguments.
The results are shown to be in excellent agreement with simulation data. 
Two collision rules are considered and their computational efficiency is 
compared.

\end{abstract}

\maketitle

\section{Introduction}

In a series of recent papers \cite{ihle_01,ihle_02a,ihle_02b}, a discrete-time 
projection operator technique was used to derive Green-Kubo relations for the 
transport coefficients of a new stochastic model---which we will call 
Stochastic Rotation Dynamics---for simulating fluctuating fluid 
dynamics \cite{male_99,male_00}.  
Explicit expressions for transport coefficients in two dimensions were derived,
and it was shown how random shifts of the collision environment could be used 
to ensure Galilean invariance for arbitrary Mach number and temperature. In 
this paper, we extend our analytical and numerical analysis to three dimensions
and consider two distinct collision rules. Expressions for the transport 
coefficients are derived and compared with simulation results.
No assumptions are made regarding molecular chaos, and the correlations 
which can develop at small mean free path are explicitly accounted for. 
The only approximation we make is to neglect fluctuations in the number 
of particles in the cells which are used to define the collision environment. 
This amounts to neglecting terms of order 
${\rm e}^{-M}$, where $M$ is the average number of particles in a cell, and 
is therefore justified in all practical calculations, where $M\geq 5$.

In the Stochastic Rotation Dynamics (SRD) algorithm, the fluid is modeled 
by particles whose position coordinates, ${\bf r}_i(t)$, and velocities,
${\bf v}_i(t)$, are continuous variables. The system is coarse-grained into 
cells of a regular lattice, and there is no restriction on the number of 
particles in a cell. The evolution of the system consists of two steps:
streaming and collision. In the streaming step, all particles are 
simultaneously propagated a distance ${\bf v}_i\tau$, where $\tau$ is the 
value of the discretized time step. For the collision step, particles are
sorted into cells, and they interact only with members of their own cell.
Typically, the simplest cell construction consisting of a hypercubic grid
with mesh size $a$ is used. As discussed in 
Refs. \cite{ihle_01} and \cite{ihle_02a}, 
a random shift of the particle coordinates before the collision step is 
required to ensure Galilean invariance. In our implementation of this  
procedure all particles are shifted by the {\em same} random vector with 
components in the interval $[-a/2,a/2]$ before the collision step. Particles 
are then shifted back to their original positions after the collision. If 
we denote the cell coordinate of the shifted particle $i$ by $\bxi_i^s$, 
the dynamics is summarized by  
\begin{eqnarray}
\label{eqm_1}
{\bf r}_i(t+\tau)&=&{\bf r}_i(t)+\tau\;{\bf v}_i(t) \\
\label{eqm_2}
{\bf v}_i(t+\tau)&=&{\bf u}[\bxi_i^s(t+\tau)]+\bomega[\bxi_i^s(t+\tau)]
\cdot\{{\bf v}_i(t)-{\bf u}[\bxi_i^s(t+\tau)]\},
\end{eqnarray}
where $\bomega(\bxi_i^s)$ denotes a stochastic rotation matrix, and 
${\bf u}(\bxi_i^s)\equiv{1\over M}\sum_{k\in\bsxi^s}{\bf v}_k$ is the 
mean velocity of the particles in cell $\bxi^s$. All particles in the 
cell are subject to the same rotation, but the rotations in different cells 
are statistically independent. There is a great deal of freedom in how the 
rotation step is implemented, and any stochastic rotation matrix consistent 
with detailed balance can be used. The dynamics of the SRD algorithm is 
explicitly constructed to conserve mass, momentum, and energy, and the 
collision process is the simplest consistent with these conservation laws. 
The algorithm is Galilean invariant, there is an $H$-theorem, and it yields 
the correct hydrodynamics equations with an ideal gas equation of state
\cite{male_99,ihle_02a}. SRD has been used to study flow around solid objects 
in both two \cite{lamu_01,lamu_02} and three \cite{alla_02} dimensions, 
dilute polymer solutions \cite{male_00a}, binary mixtures 
\cite{male_00b,hash_00}, amphiphilic mixtures \cite{saka_00,saka_02a,saka_02b},
colloids \cite{inou_01,inou_02}, and cluster structure and dynamics 
\cite{lee_01}. 

In two dimensions, the stochastic rotation matrix, $\bomega$, is typically 
taken to be a rotation by an angle $\pm\alpha$, with probability $1/2$.   
Analytic expressions for the transport coefficients in this case were 
derived in 
Refs. \cite{ihle_01,ihle_02a,ihle_02b} and shown to be in excellent agreement 
with simulation results. In three dimensions, one collision rule that has been 
discussed in the literature \cite{male_99,male_00,alla_02} consists of 
rotations by an angle $\alpha$ about a randomly chosen direction. All 
orientations of the random axis occur with equal probability. Note that 
rotations by an angle $-\alpha$ do not need to be considered, since this 
amounts to a rotation by an angle $\alpha$ about an axis with the opposite 
orientation. The viscosity of this model has been measured using a Poiseuille 
flow geometry in Ref. \cite{alla_02}. Analytical expressions for the transport 
coefficients in this case are only available in the limit of large mean free 
path, $\lambda/a\rightarrow \infty$, and for one rotation angle, 
$\alpha=90^\circ$ \cite{male_00}. In the following, we  
will refer to this collision rule as Model A. 
Another, computationally simpler collision rule, which we will refer to as 
Model B, involves rotations about one of three orthogonal rotation 
axes. In the implementation considered here, we take these to be  
$x$-, $y$- and $z$-axes of a cartesian coordinate system. At each collision 
step one of these three axes is chosen at random, and a rotation by an 
angle $\pm\alpha$ is then performed, where the sign is chosen at random. 
This procedure is fast and easy to 
implement; furthermore, only six different rotation matrices are needed, 
which are sparse and contain fixed elements of $1$, $\pm\sin(\alpha)$, 
and $\cos(\alpha)$. Our simulations have 
shown that both collision rules lead to a rapid relaxation to thermal 
equilibrium characterized by the Maxwell-Boltzmann velocity distribution.  

The outline of the  paper is as follows. In Sec. \ref{sec:hydro} we 
briefly summarize the Green-Kubo relations for the transport coefficients.
Secs. \ref{sec:random} and \ref{sec:reduced} contain detailed descriptions 
of the two collision rules, as well as analytical and numerical calculations of 
the shear viscosity, thermal diffusivity and self-diffusion coefficient at both 
large and small mean free path. 
% In Sec. \ref{sec:relax} we will discuss the relaxation behavior of both 
% methods and obtain transport coefficients in the $M\rightarrow \infty$ 
% limit. 
The work is summarized in Sec. \ref{sec:CONC}.

\section{Hydrodynamics} \label{sec:hydro}

The transport coefficients of a simple liquid are the kinematic 
shear and bulk viscosities, $\nu$ and $\gamma$, and the thermal diffusivity 
coefficient, $D_T$. Explicit expressions for the asymptotic (long-time limit) 
shear and bulk viscosities and thermal diffusivity of the SRD algorithm 
were derived in Ref. \cite{ihle_02a} using a projection operator technique. 
In particular, it was shown that the kinematic viscosities can be expressed 
in terms of the reduced fluxes in $k$-space as  
\begin{equation}\label{visc} 
\nu\left(\delta_{\beta\varepsilon}+{{d-2}\over d}{k_\beta k_\varepsilon\over 
k^2}\right) + \gamma {k_\beta k_\varepsilon\over k^2} = 
{\tau\over Nk_BT}\left.\sum_{t=0}^\infty\right.' 
\langle I_{1+\beta}(\hat\bfk,0) \vert I_{1+\varepsilon}(\hat\bfk,t)\rangle, 
\end{equation}
\noindent
while the thermal diffusivity is given by  
\begin{equation}\label{tc} 
D_T = {\tau\over c_pNk_BT^2} \left.\sum_{t=0}^\infty\right.' 
\langle I_{d+2}(\hat\bfk,0)\vert I_{d+2}(\hat\bfk,t)\rangle\;\;,
\end{equation}  
where $d$ is the spatial dimension, $c_p=(d+2)k_B/2$ is the specific heat 
per particle at constant 
pressure, and the prime on the sum indicates that the $t=0$ term has 
the relative weight $1/2$. Here and in the following we have set the particle 
mass equal to one. The thermal conductivity, $\kappa$, is related to $D_T$ by   
\begin{equation}
\kappa=\rho c_p D_T.
\end{equation} 
\noindent
The reduced fluxes in Eqs. (\ref{visc}) and (\ref{tc}) are (see Refs. 
\cite{ihle_02a,ihle_02b} for details)
\begin{equation}\label{i1+a}
I_{1+\beta}(\hat\bfk,t) = \frac{1}{\tau}\sum_i\bigg(-[v_{i\beta}(t)
\hat\bfk\cdot \Delta\bxi_i(t)+\Delta v_{i\beta}(t) 
\hat\bfk\cdot\Delta\bxi^s_i(t)]    
+{\tau \hat k_\beta\over d} v_i^2(t)   \bigg), \label{fluxvisco}
\end{equation} 
for $\beta = 1,\dots,d$, and 
\begin{eqnarray}\label{d+2}  
I_{d+2}(\hat\bfk,t) &=& \frac{1}{\tau}\sum_i\bigg(-[(v_i^2(t)/2-c_vT)
\hat\bfk\cdot \Delta\bxi_i(t)  \nonumber \\ 
&+& {1\over2}\Delta v_i^2(t) \hat\bfk\cdot \Delta\bxi^s_i(t)]   
+\tau k_BT\, \hat\bfk\cdot{\bf v}_i(t)\bigg),  \label{fluxcond}
\end{eqnarray} 
where $c_v=d\,k_B/2$ is the specific heat per particle at constant volume 
of an ideal gas, $\Delta v_j^2=v_j^2(t+\tau) -v_j^2(t)$,
$\Delta {\bxi}_i (t)= \bxi_i(t+\tau) - \bxi_i(t)$, and 
$\Delta {\bxi}^s_i (t)=\bxi_i(t+\tau) - \bxi^s_i(t+\tau)$, where 
$\bxi_i(t+\tau)$ is the cell cooridinate of particle $i$ at time $t+\tau$ 
and $\bxi^s_i(t+\tau)$ is the corresponding shifted particle cell coordinate. 
Since $\Delta\bfr_i(t) = \tau{\bf v}_j(t)$, $I_1(\hat\bfk,t)=0$ to this order 
in $k$. 

\subsubsection{Shear viscosity}

In three dimensions, the shear viscosity can be obtained if, for example,  
one takes $\hat\bfk$ in the $z$-direction and $\beta=\epsilon=1$, in the 
Green-Kubo relation, Eq. (\ref{visc}), so that 
\begin{equation}\label{sv} 
\nu = {\tau\over Nk_BT}\left.\sum_{t=0}^\infty\right.' \langle I_2(\zhat,0)
\vert I_2(\zhat,t)\rangle. 
\end{equation}
There are two contributions to the reduced fluxes, namely kinetic and 
rotational, so that 
\begin{equation}
I_2(\zhat,t)= I_2^{kin}(\zhat,t) + I_2^{rot}(\zhat,t) \label{I2}
\end{equation} 
where
\begin{equation}
I_2^{kin}(\zhat,t) = -\frac{1}{\tau}\sum_i v_{ix}(t) \Delta\xi_{iz}(t) 
\end{equation} 
and 
\begin{equation}
I_2^{rot}(\zhat,t) = -\frac{1}{\tau}\sum_i \Delta v_{ix}(t) 
\Delta\xi^s_{iz}(t).
\end{equation} 
Contributions to $I_2^{kin}$ come from the streaming step, whereas the 
collisions and shifts contribute to $I_2^{rot}$. There are corresponding 
kinetic, rotational, and mixed contributions to the shear viscosity. 

\subsubsection{Thermal diffusivity}

Similarly, setting $d=3$ and taking $\hat\bfk$ in the $z$-direction in Eq. 
(\ref{tc}), one has 
\begin{equation}\label{dt0} 
D_T = {\tau\over c_pNk_BT^2} \left.\sum_{t=0}^\infty\right. ' \langle I_5(\zhat,0) \vert I_5(\zhat, t)\rangle \;\;.
\end{equation}
\noindent 	
Again, the reduced flux can be divided into the kinetic and rotational 
contributions, so that   
\begin{equation}
I_5(\zhat,t)= I_5^{kin}(\zhat,t) + I_5^{rot}(\zhat,t), \label{I5}
\end{equation} 
where
\begin{equation}\label{I5_k}  
I_5^{kin}(\zhat,t) = \frac{1}{\tau}\sum_i \left\{\left( c_vT - 
\frac{v_i^2(t)}{2}\right)\Delta\xi_{iz}(t) +\tau k_BT\, v_{iz}(t)\right\} 
\end{equation} 
and 
\begin{equation}\label{I5_r}   
I_5^{rot}(\zhat,t) = -\frac{1}{\tau}\sum_i {1\over2}\Delta v_j^2(t) 
\Delta\xi^s_{iz}(t).   
\end{equation} 

\section{Model A: rotation around a random axis} \label{sec:random}

\begin{figure}
\begin{center}
\leavevmode
\includegraphics[width=4in]{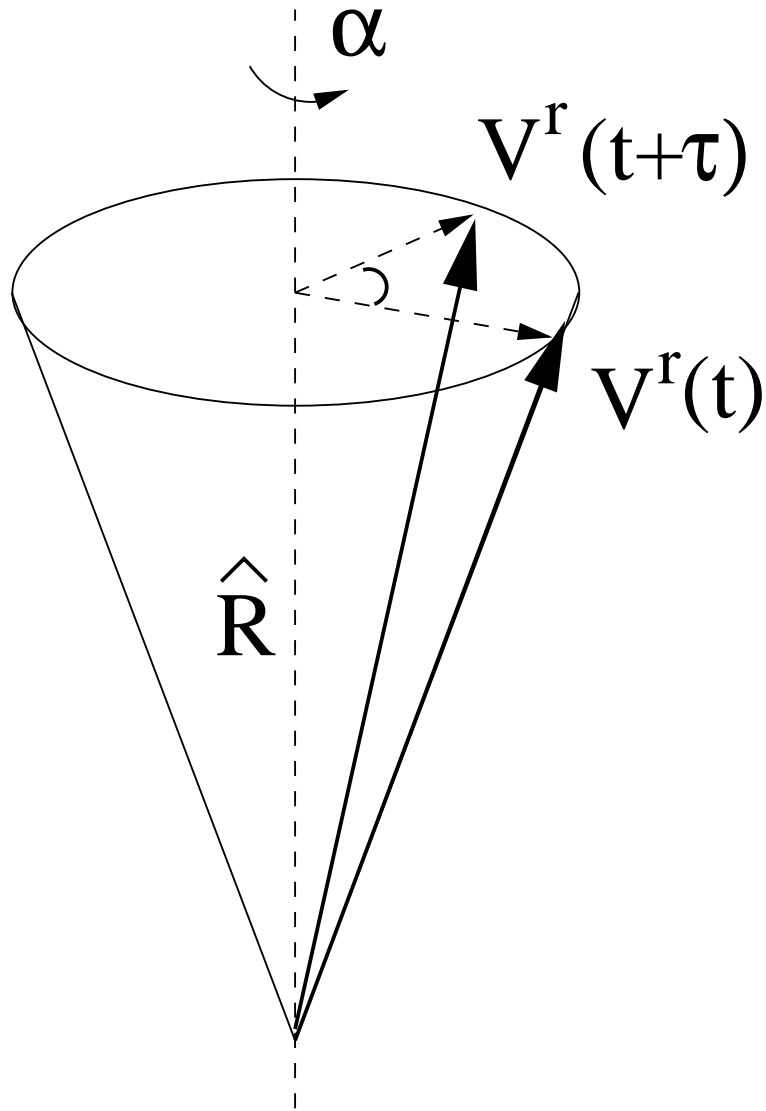}
\caption{Rotation of the vector $\bfv^r$ around a random direction 
$\hat{\bf R}$ by the angle $\alpha$.}\label{rotation}
\end{center}
\end{figure}

As discussed in the introduction, one choice of collision rule is a rotation 
by an angle $\alpha$ about a randomly chosen axis (see Fig. \ref{rotation}). 
This collision rule has been used in a recent study of Poiseuille flow and  
flow around a spherical obstacle, and was shown to yield excellent results
\cite{alla_02}. Denote the random vector by $\bf {\hat{R}}$; the 
post-collision velocity of a particle at time step $t+\tau$ can then be 
written as
\begin{equation}
{\bf v}(t+\tau)={\bf u}_{{\bsxi}^s}(t)+{\bf v}_{\perp}^{r}(t) \cos(\alpha)+
({\bf{v}}_{\perp}^r (t)\times {\bf{\hat{R}}}) \sin(\alpha) + 
{\bf v}_{\parallel}(t) \label{evol}
\end{equation} 
where $\perp$ and $\parallel$ denote the components of a vector that are 
perpendicular and parallel to the random axis ${\bf \hat{R}}$; the relative 
velocity ${\bf v}^r (t)={\bf v}(t) - {\bf u_{\bsxi^s}}(t)$.  

\begin{figure}
\begin{center}
\leavevmode
\includegraphics[width=5in,angle=270]{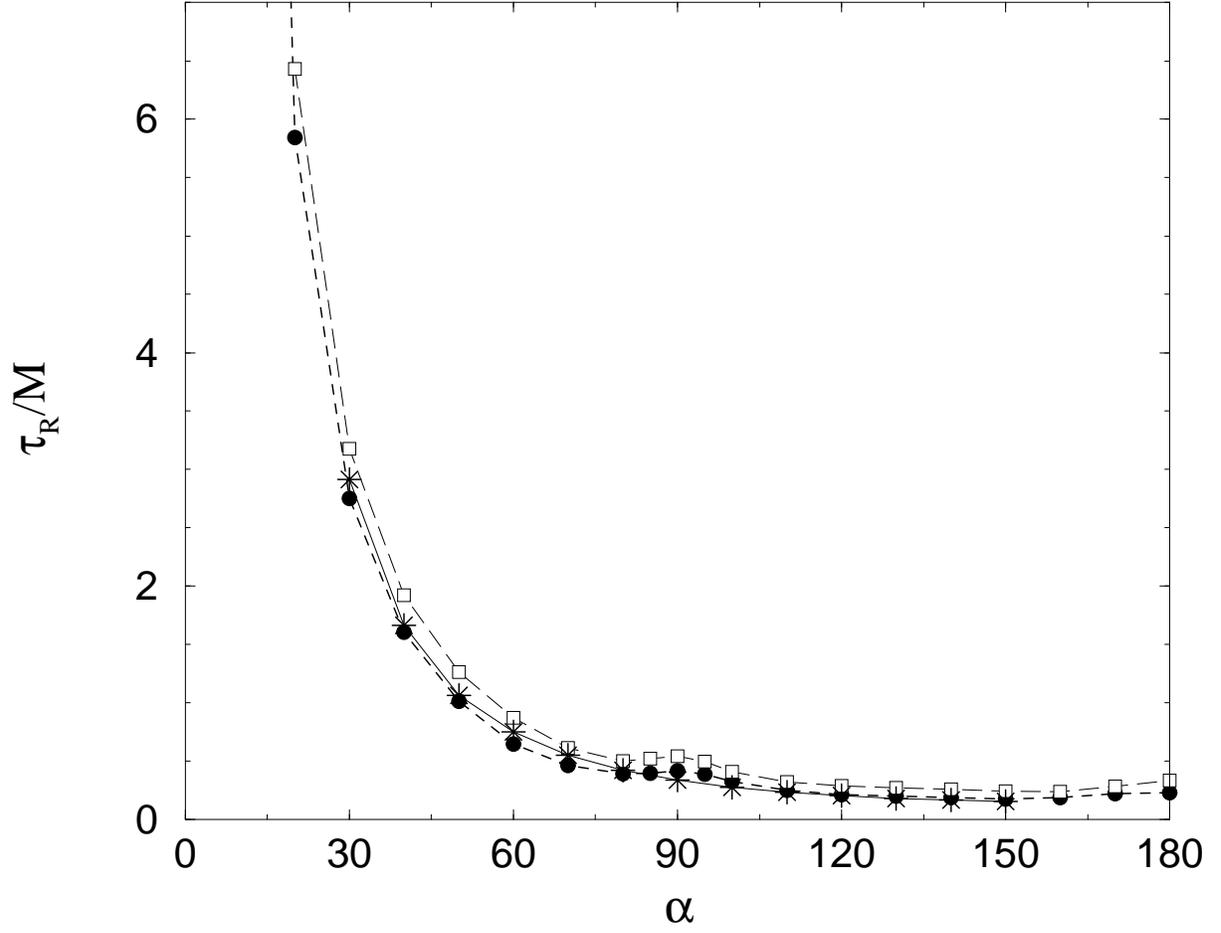}
\caption{The normalized relaxation time, $\tau_R/M$, of the fourth moment of 
the velocity distribution, $S_4=\sum_i(v_{ix}^4+v_{iy}^4+v_{iz}^4)$ as a 
function of the rotation angle $\alpha$ for $M=20$, where $M$ is the average 
number of particles per cell. The data for Model A
($\ast$) were obtained for $\lambda/a=1.15$, while the data for Model B 
correspond to $\lambda/a=1.15$ ($\lbullet$) and $\lambda/a=0.0361$ ($\ssquare$). 
Parameters: $L/a=32$ and $\tau=1$. }\label{PRE_3DW_5}
\end{center}
\end{figure}

The relaxation to thermal equilibrium is characterized by the decay rate of 
the $H$-function \cite{ihle_02a}. However, a simpler procedure is to monitor 
the relaxation of the fourth moment, 
$S_4=\sum_i\,(v_{ix}^4+v_{iy}^4+v_{iz}^4)$, 
of the velocity distribution. This was done in Ref. \cite{ihle_02a} in two 
dimensions, where it was shown that $S_4$ relaxes exponentially to
the equilibrium value given by the Maxwell-Boltzmann distribution
with a relaxation time, $\tau_R$, which is essentially temperature independent. 
Furthermore, it was found that $\tau_R$ is proportional to the average number of
particles in a cell, $M$, and depends strongly on the value of the rotation
angle $\alpha$. It diverges approximately as $\tau_R\sim \alpha^{-2}$ for
$\alpha \rightarrow 0$, since there are no collisions in this limit, and thermal
equilibrium cannot be achieved. As can be seen in Fig. \ref{PRE_3DW_5}, 
similar behavior is observed in three dimensions for both Models A and B. 

\subsection{Large mean free path approximation}\label{sec:Almfp} 

\subsubsection{Shear viscosity}\label{sec:Asv} 

For large mean free path, $\lambda/a\rightarrow \infty$, the rotational 
contributions to the reduced flux, $I_2^{rot}(\zhat,t)$, in Eq. (\ref{I2}) 
can be neglected, so that the shear viscosity can be expressed as 
\begin{equation} \label{sv2}
\nu = {\tau\over Nk_BT}\left.\sum_{n=0}^\infty\right.' C_n \;,
\end{equation}
where 
\begin{equation}\label{defcn}   
C_n\equiv\langle I_2^{kin}(\zhat,0) \vert I_2^{kin}(\zhat,n\tau)\rangle = 
{1\over\tau^2}\sum_{ij}\,\langle v_{ix}(0)\Delta\xi_{iz}(0)v_{jx}(n\tau)
\Delta\xi_{jz}(n\tau) \rangle. 
\end{equation}  
As discussed in Ref. \cite{ihle_02b}, except for the $t=0$ contribution, 
$C_0$, it is a good approximation to replace $\Delta\xi_{iz}$ by $\tau v_{iz}$ 
when evaluating $C_n$. In the following, we therefore first evaluate $\nu$ 
using this approximation, and then discuss the required correction term.  

The relevant components of Eq. (\ref{evol}) can be written as
\begin{eqnarray} 
\label{evolvx} 
v_{ix}(t+\tau)&=&u_{\xi x}(t)+c[v_{ix}(t)-u_{\xi x}(t)-\sum_{\beta} 
(v_{i \beta}(t)-u_{\xi \beta}(t))R_\beta R_x]\\ \nonumber
& & \ \ \ +s[(v_{iy}(t)-u_{\xi y}(t))R_z-(v_{iz}(t)-u_{\xi z}(t))R_y] 
\\ \nonumber	
& & \ \ \ +\sum_{\beta} (v_{i \beta}(t)-u_{\xi \beta}(t))R_\beta R_x ,  
\\  
\label{evolvz} 
v_{iz}(t+\tau)&=&u_{\xi z}(t)+c[v_{iz}(t)-u_{\xi z}(t)-\sum_{\beta} 
(v_{i \beta}(t)-u_{\xi \beta}(t))R_\beta R_z]\\ \nonumber
& & \ \ \ +s[(v_{ix}(t)-u_{\xi x}(t))R_y-(v_{iy}(t)-u_{\xi y}(t))R_x] 
\\ \nonumber	
& & \ \ \ +\sum_{\beta} (v_{i \beta}(t)-u_{\xi \beta}(t))R_\beta R_z ,  
\end{eqnarray} 
where $c=\cos(\alpha)$, $s=\sin(\alpha)$, 
${\bf u}_\xi={1\over M}\sum_{k\in\xi} {\bf v}_k$, and the sum runs over all 
particles in the cell occupied by particle $i$ at $t=n\tau$. Defining  
\begin{equation} 
A_n = \sum_{ij}\,\langle v_{ix}(0)v_{iz}(0)v_{jx}(n\tau)v_{jz}(n\tau) \rangle, 
\end{equation}  
we have 
\begin{equation}
A_0=\sum_{ij}\langle v_{ix}v_{iz} v_{jx}v_{jz} \rangle = 
    \sum_i\langle v_{ix}v_{iz} v_{ix}v_{iz} \rangle = N(k_BT)^2,  
\end{equation} 
so that there are only contributions from $j=i$. The second term in the series 
is
\begin{equation} 
A_1=\sum_{ij}\langle v_{ix}v_{iz}v_{jx}(\tau)v_{jz}(\tau)\rangle,    
\end{equation} 
where $v_{jx}(\tau)$ and $v_{jz}(\tau)$ are given by Eqs. (\ref{evolvx}) 
and (\ref{evolvz}), respectively. There are both diagonal ($j=i)$ and 
off-diagonal $(j\ne i)$ contributions to $A_1$. Making use of the following 
averages over the random vector ${\bf \hat{R}}$, 
\begin{equation}
\langle R_\beta^2 \rangle = 1/3  
\end{equation} 
and 
\begin{equation}
\langle R_\beta^2 R_{\beta^\prime}^2 \rangle = 
1/15 + 2/15\ \delta_{\beta,\beta^\prime} \;,
\end{equation} 
the diagonal contribution is found to be 
\begin{equation}
\langle v_{ix}v_{iz}v_{ix}(\tau)v_{iz}(\tau)\rangle = (k_BT)^2 \zeta_A, 
\end{equation} 
where 
\begin{equation}
\zeta_A = \frac{1}{3}\bigg[  
\big[c^2-s^2+{2\over 5} (c-1)^2 \big] \big( {1\over M}-1 \big)^2+2c
\big(1-{1\over M^2}\big)+{1\over M}\big(2+{1\over M}\big) \bigg] \;.
\end{equation} 
\noindent
The off-diagonal contribution comes from particles $j$ which are in the same 
cell as particle $i$ at $t=0$. This contribution is equal to  
\begin{equation}
\langle v_{ix}v_{iz}v_{jx}(\tau)v_{iz}(\tau)\rangle = \eta_A,  
\end{equation} 
\noindent
where
\begin{equation}
\eta_A = \frac{2}{15M^2} (6c-1)(c-1) \;\;.
\end{equation} 
Since there are $M-1$ off-diagonal contributions, it follows that   
\begin{equation} 
A_1 = N(k_BT)^2[\zeta_A+(M-1)\eta_A]. 
\end{equation}  

\begin{figure}
\begin{center}
\leavevmode
\includegraphics[width=3.5in,angle=270]{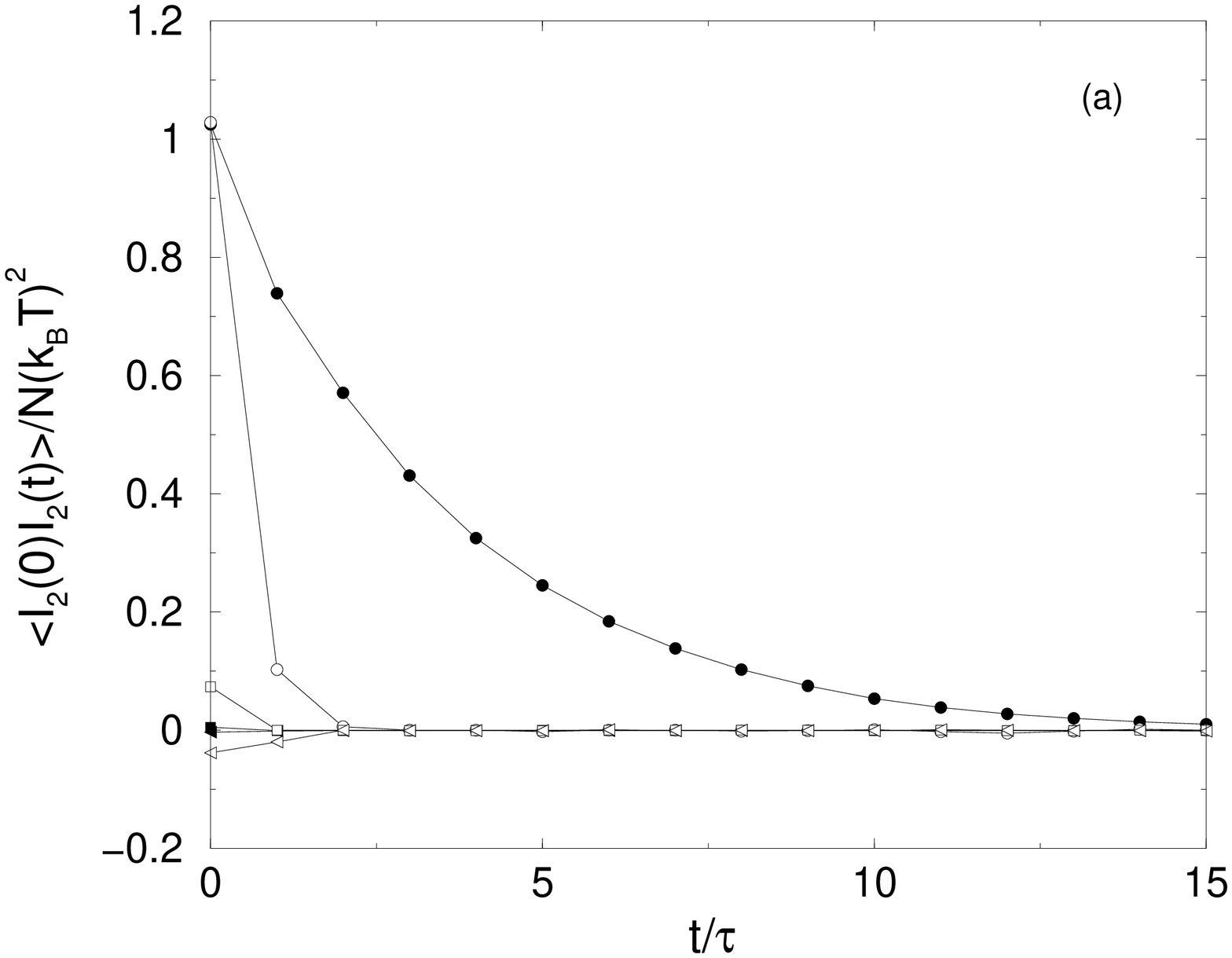}
\bigskip
\includegraphics[width=3.5in,angle=270]{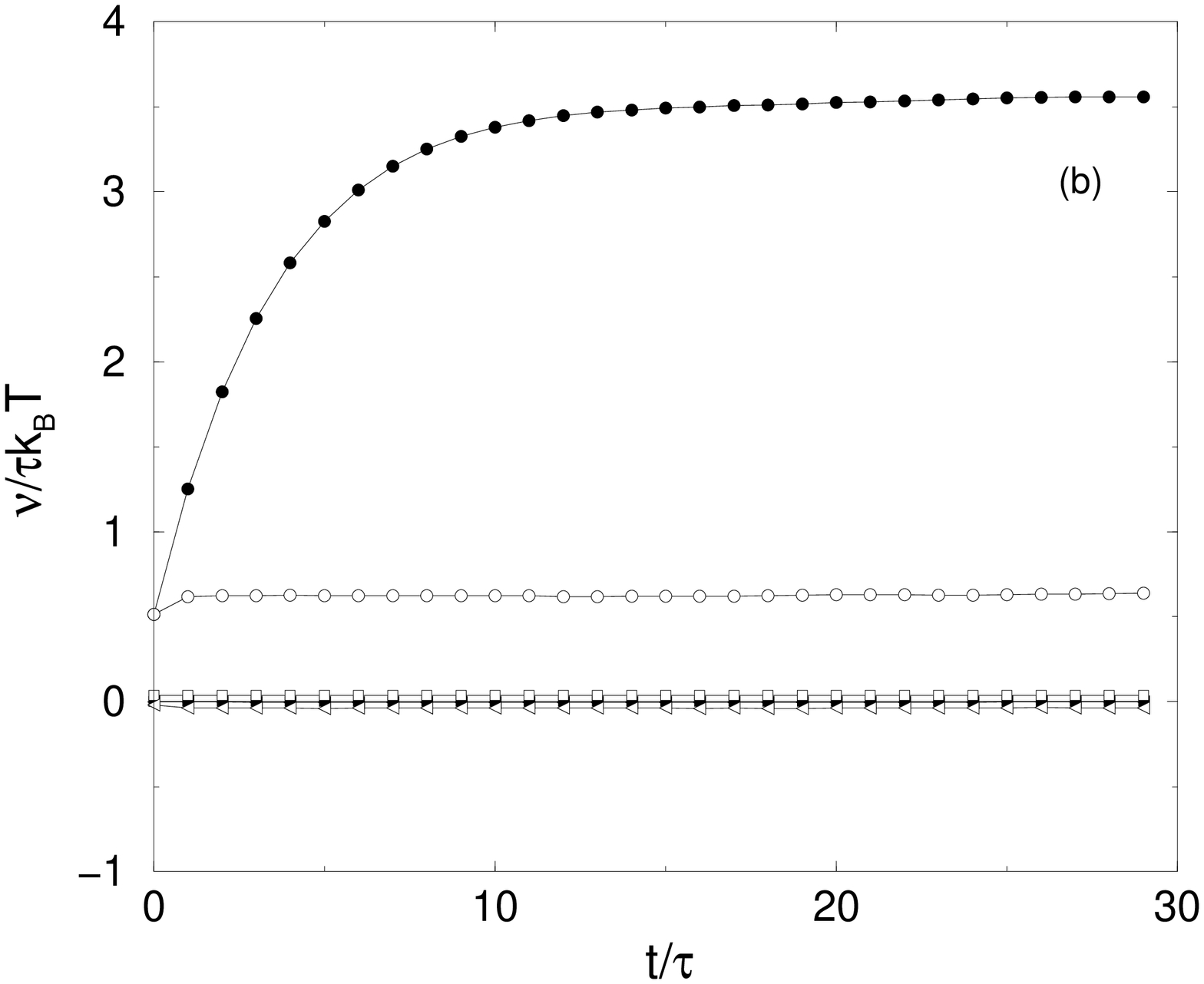}
\end{center}
\caption{(a) Normalized correlation functions $\langle I_2(0)I_2(t)\rangle
/N(k_BT)^2$ for Model A as a function of time for $\alpha=30^\circ$ 
(solid symbols) 
and $\alpha=150^\circ$ (unfilled symbols). For $\alpha=30^\circ$, the kinetic, 
rotational, and mixed contributions are indicated by $\lbullet$, 
$\sblacksquare$, and $\blacktriangleleft$, respectively. For $\alpha=150^\circ$,
the kinetic, rotational, and mixed contributions are indicated by  
$\lcirc$, $\ssquare$, and $\lhd$, respectively.  
(b) Normalized time dependent kinematic viscosity, $\nu(t)/\tau k_BT$. 
Symbols are the same as in part (a). 
Parameters: $L/a=32$, $\lambda/a=2.309$, $\tau=1$, and $M=20$.
The data were obtained by time averaging over 75000 iterations.
}
\label{visco_corr}
\end{figure}

\begin{figure}
\begin{center}
\leavevmode
\includegraphics[width=3.5in,angle=270]{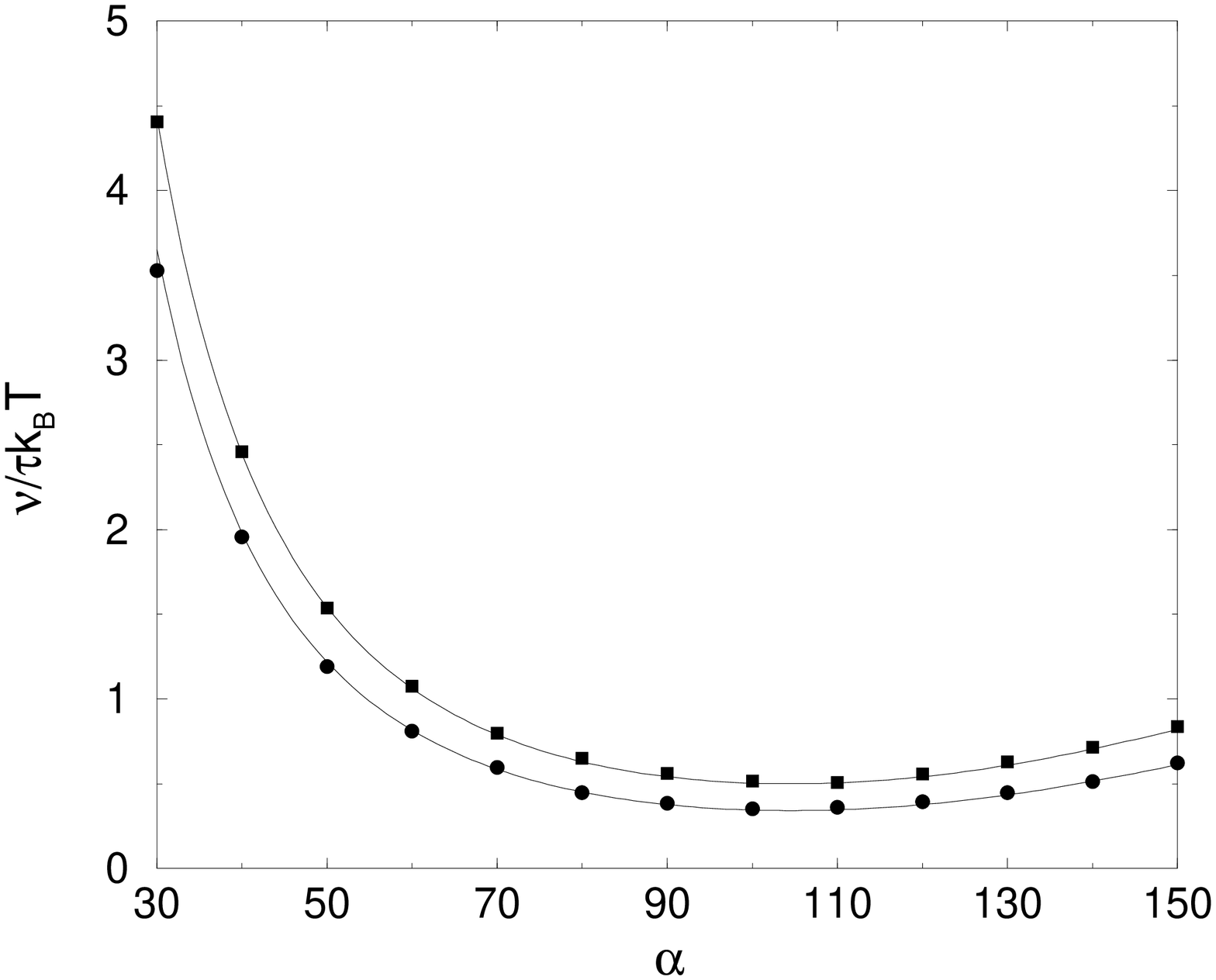}
\bigskip 
\includegraphics[width=3.5in,angle=270]{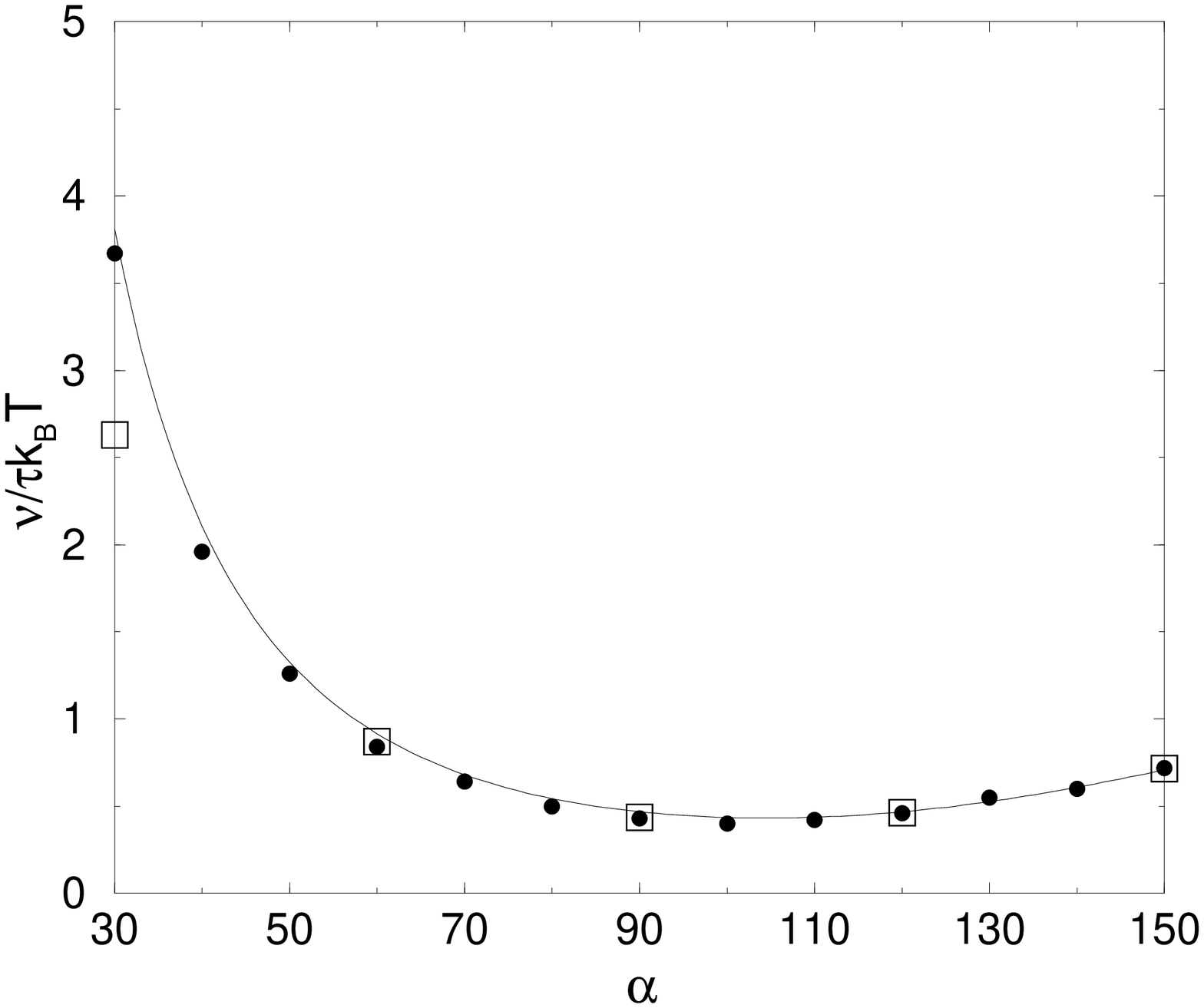}
\caption{Normalized kinematic viscosity, $\nu/\tau k_BT$, for Model A as 
a function of 
the collision angle $\alpha$. (a) Data for $L/a=32$, $\lambda/a = 2.309$, 
$\tau=1$, and  $M=5$ ($\sblacksquare$) and $M=20$ ($\lbullet$). (b) Data 
for $L/a=32$, $\lambda/a = 1.02$, $\tau=1$, and $M=20$. The bullets are 
results obtained using the Green-Kubo relation, and the unfilled boxes 
($\ssquare$)  
are data for the kinematic viscosity obtained in Ref. \cite{alla_02} by 
fitting the one-dimensional velocity profile of forced flow between parallel 
plates. The lines are the theoretical prediction, Eq. (\ref{kin_vis_final}), 
for the corresponding parameter values. 
The data were obtained by time averaging over 75000 iterations.
The deviation of the data point $\ssquare$ at $\alpha=30^\circ$ is due to 
finite Knudsen number effects.   
}\label{viscolarge}
\end{center}
\end{figure}

\begin{figure}
\begin{center}
\leavevmode
\includegraphics[width=3.5in,angle=270]{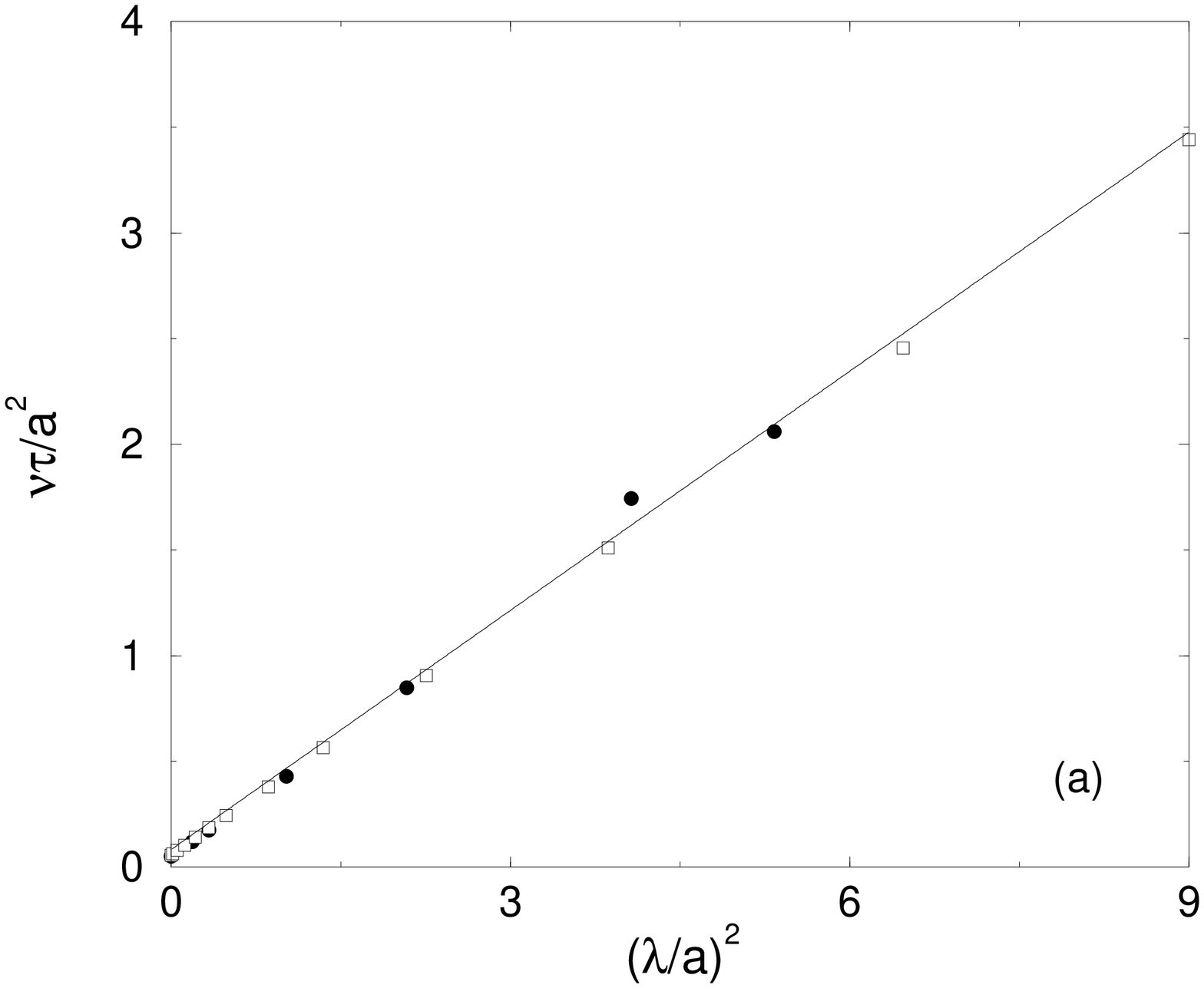}
\bigskip 
\includegraphics[width=3.5in,angle=270]{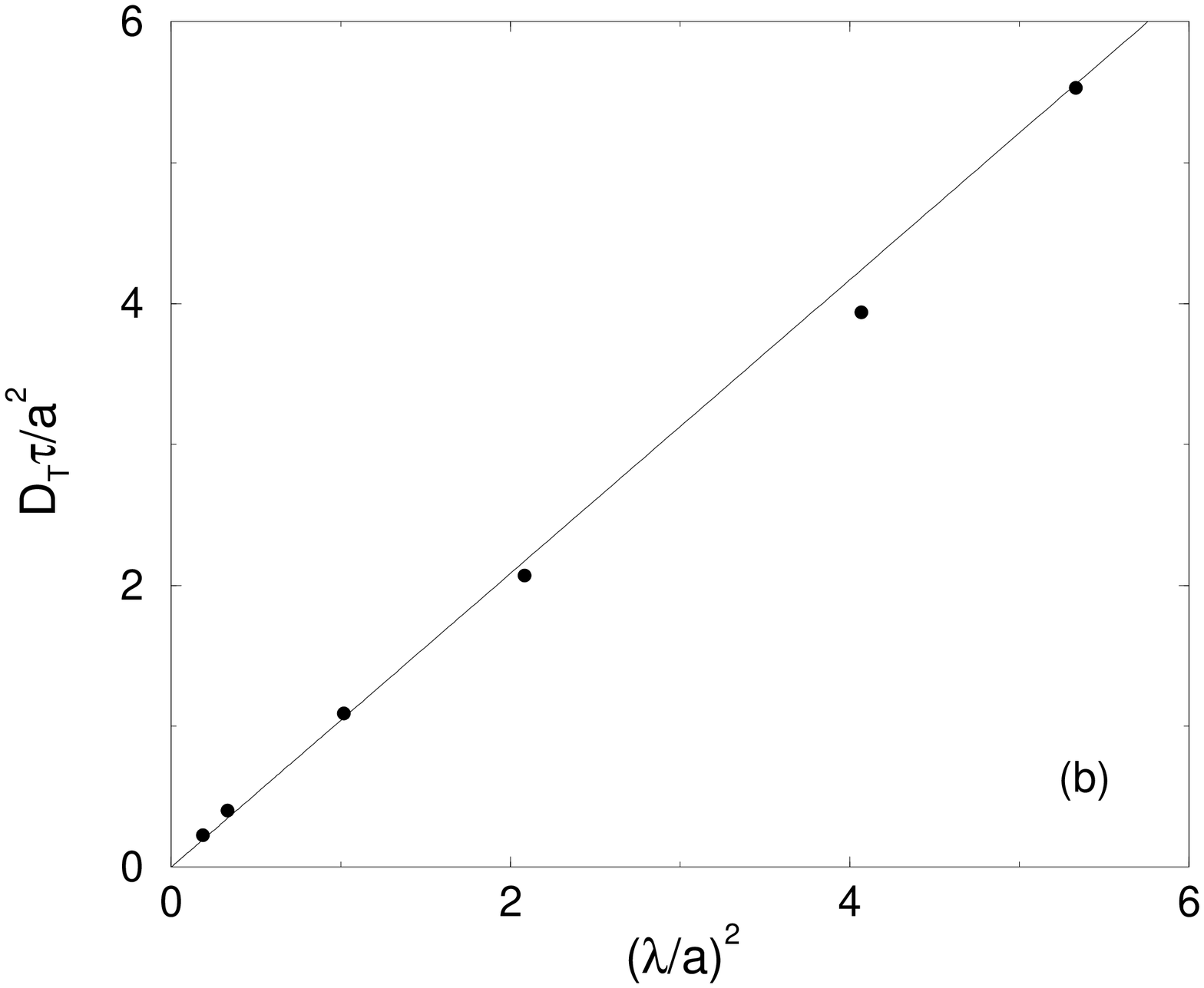}
\caption{a) Normalized kinematic viscosity, $\nu\tau/a^2$, and b) thermal 
diffusivity, $D_T\tau/a^2$, for Model A as functions of $(\lambda/a)^2$ 
for collision angle $\alpha=90^\circ$. 
The bullets are data obtained using Green-Kubo relations. The unfilled 
boxes ($\ssquare$) are data for the kinematic viscosity obtained in Ref. 
\cite{alla_02} by fitting the one-dimensional velocity profile of 
forced flow between parallel plates. The solid line is the theoretical 
prediction, Eqs. (\ref{kin_vis_final}) and (\ref{cond_final}).
Parameters: $L/a=32$, $\tau=1$, and $M=20$. 
The data were obtained by time averaging over 75000 iterations.
}\label{alpha90}
\end{center}
\end{figure}

The behavior over longer time intervals can be analyzed in a similar fashion. 
Consider $A_2$. Following the arguments of the last paragraph, there is a 
diagonal contribution proportional to $\zeta_A^2$, and an off-diagonal 
contribution proportional to $2(M-1)\eta_A\zeta_A$, since at each time step, 
$M-1$ particles become correlated with particle $i$, and particle $j$ can 
become correlated with particle $i$ at either of the two time steps.
Note, however, there are now additional---higher order---contributions which
arise, for example, when particle $j$ becomes correlated with particle $k$ 
which then becomes correlated with particle $i$. It is easy to see that 
these contributions carry additional factors of $1/M$ and are thus of higher 
order than the diagonal and direct off-diagonal contributions considered above.
However, these higher off-diagonal contributions can be summed in the 
geometric series
\begin{equation}\label{cn}
A_n/N(k_BT)^2 = [\zeta_A+(M-1)\eta_A]^n\approx
\zeta_A^n+n(M-1)\eta_A\zeta_A^{n-1}+\cdots,
\end{equation}                                       
so that 
\begin{eqnarray}   
\label{kin_vis}
\nu &=& k_BT\,\tau\left({1\over2} + \sum_{j=1}^\infty[\zeta_A+(M-1)\eta_A]^j
\right) \nonumber\\    
&=&{k_BT\,\tau\over2}\left( \frac{5}{(1-{1\over M})
[2-\cos(\alpha)-\cos(2\alpha)]}-1 \right).  
	\end{eqnarray}  
As discussed above, there is an additional finite cell size 
correction to this result. It arises from the fact that the substitution 
$\Delta\xi_{iy}=\tau v_{iy}$ in $C_0$ is not precisely correct. Rather, 
it can be shown that \cite{ihle_02b} 
\begin{equation}\label{corr}  
C_0 \approx A_0 +  N{k_BT\over6}{a\over\tau}^2 =    
N(k_BT)^2\left[1+{1\over6}(a/\lambda)^2\right] 
\end{equation} 
for $\lambda\gg a$. Using this result in (\ref{sv2}), the corrected kinematic 
viscosity is
\begin{equation}
\nu = {k_BT\,\tau\over2}\left( \frac{5}{(1-{1\over M})
[2-\cos(\alpha)-\cos(2\alpha)]}-1 \right) + \frac{a^2}{12 \tau} 
\;. \label{kin_vis_final}
\end{equation}
Note that although this additional term is in general negligibly small 
in three dimensions, it can dominate the viscosity in two dimensions 
\cite{ihle_02b}. In particular, for $M\to\infty$, the viscosity in 
Model A takes the 
minimum value 
\begin{equation}\label{amin}   
\nu_{min}^A = \tau k_BT\left[{3\over10} + {1\over12}\left({a\over\lambda}
\right)^2\right]
\end{equation} 
at $\alpha\approx104.48^\circ$. In contrast, in two dimensions, the minimum 
is at $\alpha=90^\circ$ for $M\to\infty$, and 
\begin{equation}\label{nu2d}
\nu_{min}^{2d} = {\tau k_BT\over12}\left({a\over\lambda}\right)^2. 
\end{equation}    
In this limit, the finite cell size correction provides the sole contribution 
to the viscosity in two dimensions. The viscosity for Model A is always 
larger than the viscosity in two dimensions.  
In order to determine the accuracy of (\ref{kin_vis_final}), 
we have performed simulations using a system of  
linear dimension $L/a=32$, using $\tau=1$, and $M=5$ and $20$ 
particles per cell. Fig. \ref{visco_corr}a contains a plot of the normalized 
correlation function $\langle I_2(0) I_2 (t)\rangle / N (k_B T)^2$
as a function of time for two different collision angles,  
$\alpha=30^\circ$ and $150^\circ$. As expected, the correlations decay much 
faster for the larger collision angle. The resulting time dependent  
kinematic viscosity is shown in Fig. \ref{visco_corr}b, and the normalized 
asymptotic value of the viscosity, $\nu/(\tau k_BT)$, is plotted in Fig. 
\ref{viscolarge}a as a function of $\alpha$ for $\lambda/a=2.309$, and $M=5$ 
and 20, and in Fig. \ref{viscolarge}b for $\lambda=1.02$ and $M=20$. 
The agreement between the analytical result and simulation 
data is excellent. Fig. \ref{alpha90}a contains a plot of the normalized 
kinematic viscosity, $\nu\tau/a^2$, as a function of $(\lambda/a)^2$ 
for $\alpha=90^\circ$ and $M=20$. 
Also shown in Fig. \ref{alpha90}a are data ($\lbullet$) for the viscosity 
obtained by fitting the one-dimensional velocity profile of forced flow 
between parallel plates in three dimensions \cite{alla_02}. Again, the 
agreement between both sets of data and theory is excellent.   

\subsubsection{Thermal diffusivity}\label{sec:tda}

As discussed in the previous section, for large mean free
path, the rotational contributions to the thermal diffusivity in Eq. 
(\ref{dt0}) can be neglected.
Furthermore, finite cell size corrections of the type discussed in the
last section do not occur in the calculation of the thermal diffusivity, 
so that $\Delta\xi_{iz}$ can be replaced by $\tau v_{iz}$. The thermal
diffusivity can therefore be expressed as
\begin{equation}\label{dt}
D_T = {\tau\over c_pNk_BT^2} \left.\sum_{n=0}^\infty\right. ' B_n \;,
\end{equation}
\noindent
where $B_n\equiv\langle I_5^{kin}(\zhat ,0) \vert I_5^{kin}(\zhat ,n\tau)
\rangle$ with  
\begin{equation}\label{i5}
I_5^{kin}(\hat z,t)=\sum_{i=1}^N \left( c_p T -\frac{v_i^2 (t)}{2}  \right) v_{iz} .
\end{equation}
Since
\begin{equation}\label{i6}
\langle c_p T (c_p T - {v_i^2 \over 2}) v_{iz} v_{jz} (n \tau) \rangle = 0
\end{equation}
for any value of $n$, it can be shown that                  
\begin{equation}
B_0= \frac{5}{2} N (k_B T)^3 \;\;.
\end{equation}
The next term in the series is 
\begin{equation}
B_1 = \frac{N}{4} \langle v_i^2 v_i^2 (\tau) v_{iz} (\tau) v_{iz} \rangle -
{Nc_p T \over 2} \langle v_i^2 (\tau) v_{iz} (\tau) v_{iz} \rangle \;. 
\label{conduct1}
\end{equation} 
In Appendix A it is shown using quaternion algebra that  
\begin{equation} \label{B1son}
B_1 = \frac{5}{2}N (k_B T)^3 \Theta , 
\end{equation} 
where 
\begin{equation} 
\Theta = \left({\gamma_1\over 3} + {\gamma_2 \over 3M} \left[1- {2 \over M}
\right] + {\gamma_3 \over 15M^3} \right)
\end{equation} 
and 
\begin{eqnarray}
\gamma_1 &=& (1+2c), \\
\gamma_2 &=& {2 \over 5} (c-1) (8c-3), \\
\gamma_3 &=& {128 \over 5} (1-c)^2 \;.
\end{eqnarray} 

\begin{figure}
\begin{center}
\leavevmode
\includegraphics[width=3.5in,angle=270]{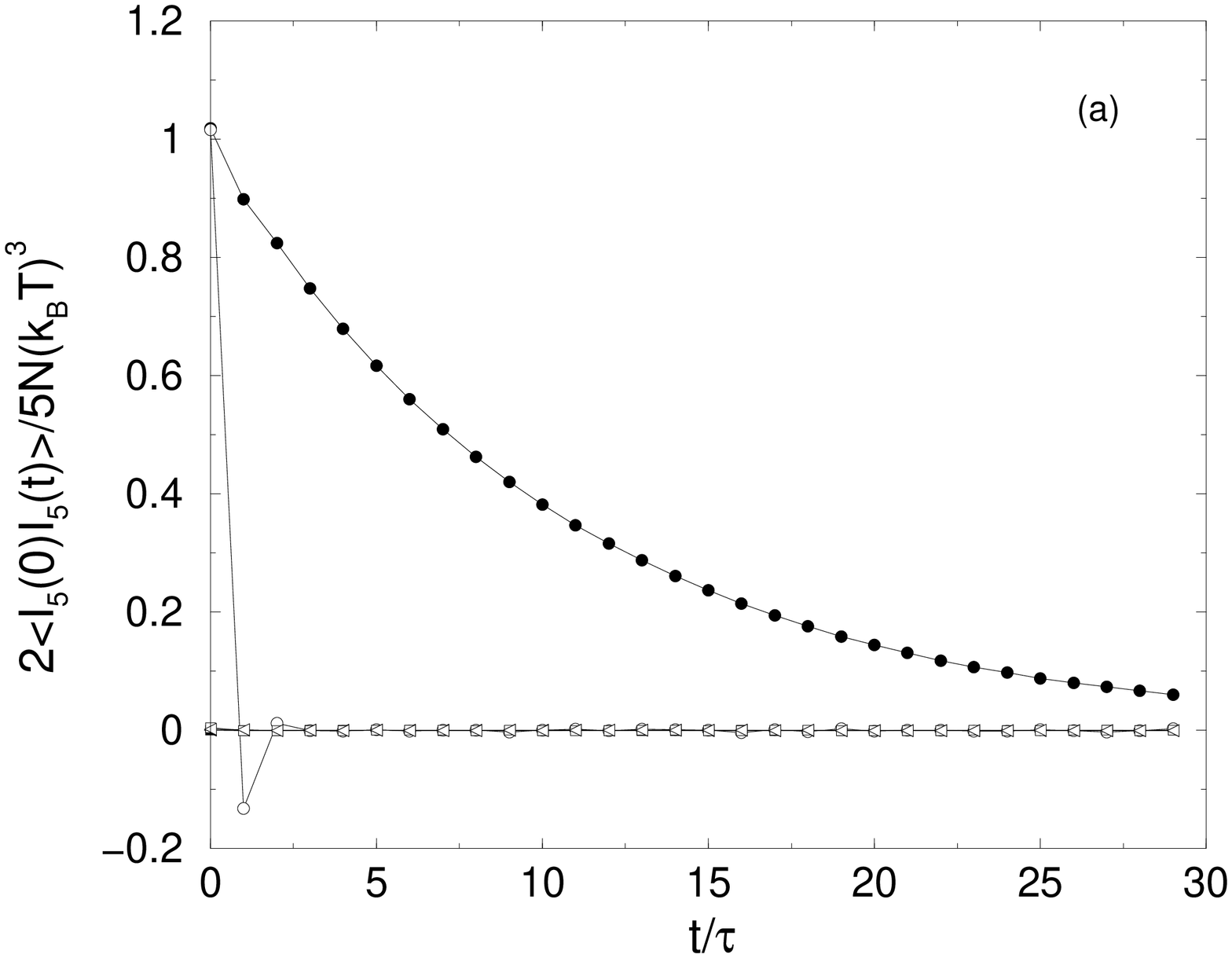}
\bigskip
\includegraphics[width=3.5in,angle=270]{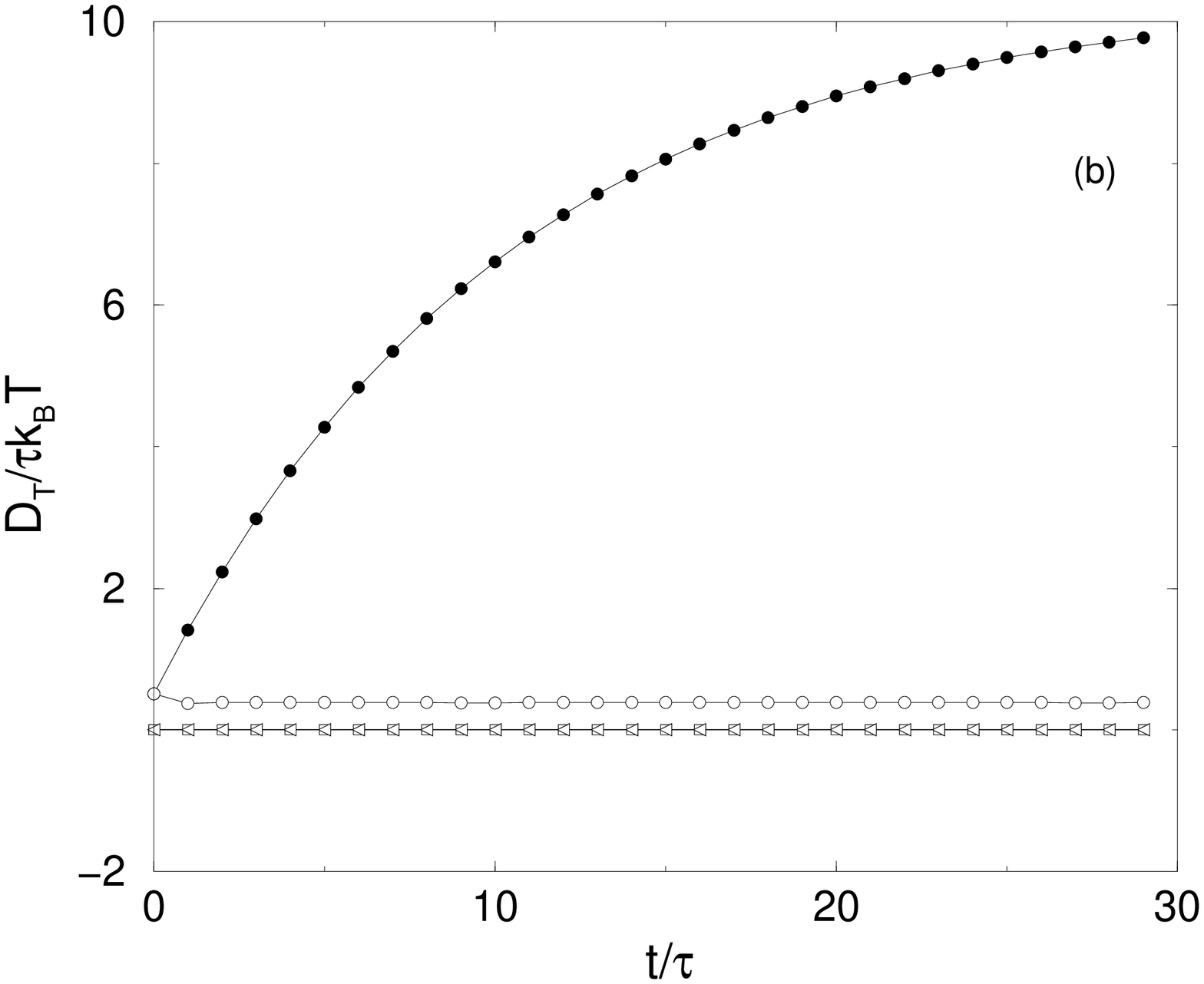}
\end{center}
\caption{(a) Normalized correlation functions $2 \langle I_5(0)I_5(t)\rangle
/5N(k_BT)^3$ for Model A as a function of time for $\alpha=30^\circ$ 
(filled symbols) 
and $\alpha=150^\circ$ (unfilled symbols). For $\alpha=30^\circ$, the kinetic, 
rotational, and mixed contributions are indicated by $\lbullet$, 
$\sblacksquare$, and $\blacktriangleleft$, respectively. For $\alpha=150^\circ$,
the kinetic, rotational, and mixed contributions are indicated by  
$\lcirc$, $\ssquare$, and $\lhd$, respectively.  
(b) Normalized time dependent thermal diffusivity, $D_T(t)/\tau k_BT$. 
Symbols are the same as in part (a). 
Parameters: $L/a=32$, $\lambda/a=2.309$, $\tau=1$, and $M=20$.
The data were obtained by time averaging over 75000 iterations.
}
\label{cond_corr}
\end{figure}

\begin{figure}
\begin{center}
\leavevmode
\includegraphics[width=5in,angle=270]{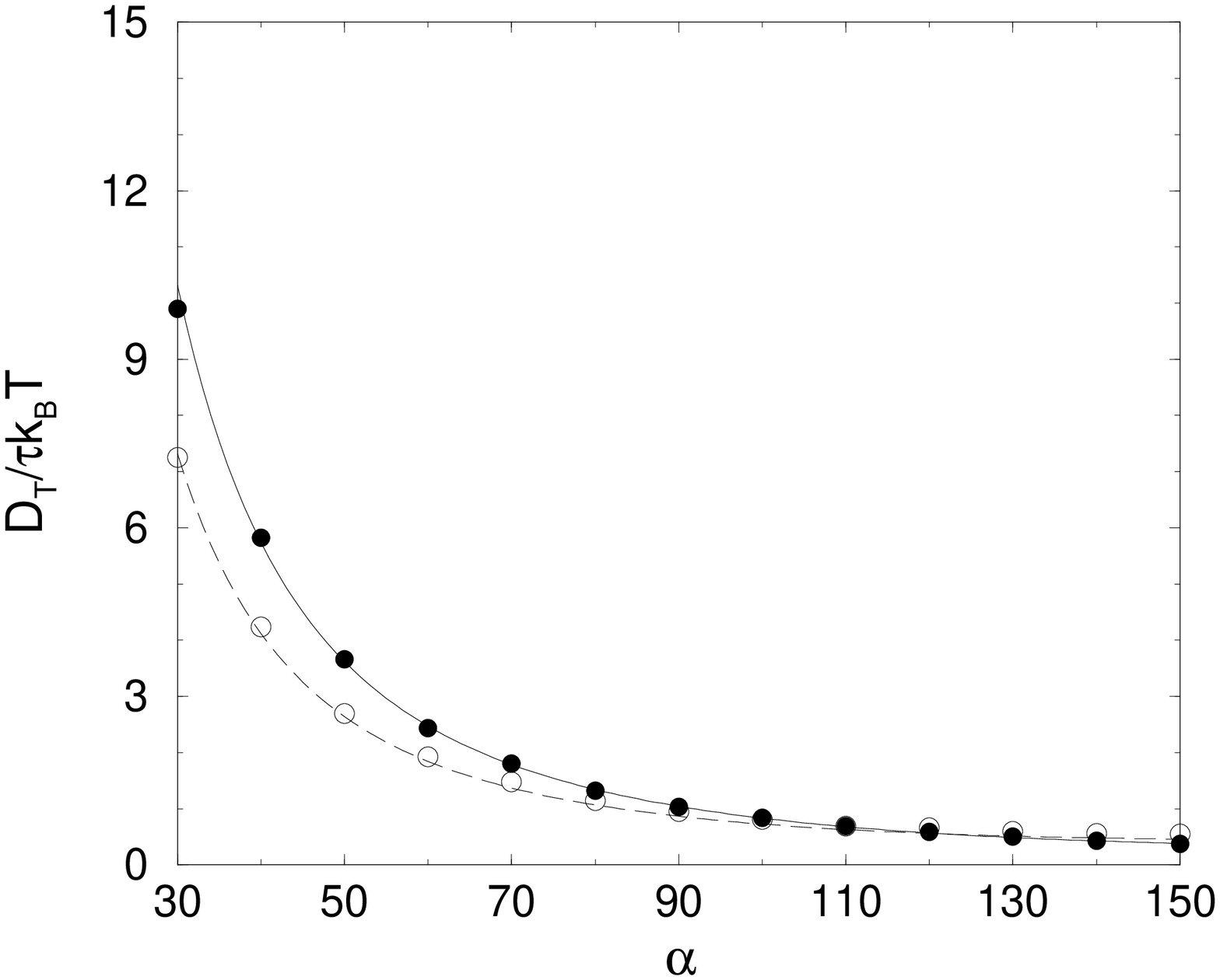}
\caption{Normalized thermal diffusivity, $D_T/\tau k_BT$, for Model A as a 
function of collision angle $\alpha$. The lines are the theoretical prediction, 
Eq. (\ref{conductivity}). The data were obtained by time averaging over 
75000 iterations. Parameters: $L/a=32$, $\lambda/a = 2.309$, 
$\tau=1$, and $M=5$ ($\lcirc$) and $M=20$ ($\lbullet$).  
}
\label{condlarge}
\end{center}
\end{figure}

Using quaternion algebra (see Appendix B), it can be shown  
in the $M \rightarrow \infty$ limit that the sum in Eq. (\ref{dt}) is a 
geometric series. Furthermore, direct calculations in two dimensions 
\cite{ihle_02b} and for Model B (see Sec. \ref{sec:lmfpb}) suggests that 
this remains true in general. Assuming this is true here, the diagonal 
contributions to the thermal diffusivity are given by   
\begin{equation} 
B_n = {5 \over 2} N(k_BT)^3 \Theta^n,   
\end{equation}  
so that carrying out the sum in Eq. (\ref{dt}), 
\begin{eqnarray}   
\label{conductivity}
D_T &=& k_BT\,\tau\left({1\over2} + \sum_{j=1}^\infty \Theta^j \right) 
\nonumber\\    
&=& { k_BT\,\tau \over 2}  \,\left( \frac{75\,M^3\,{\csc^2(\alpha/2)}} 
{2\,\left[ -64 + 5\,M\, \left( 6 + M\,\left[ -3 + 5\,M \right]  \right)  + 
 8\,\left( 8 + 5\,\left[ -2 + M \right] \,M \right) \, \cos(\alpha)\right]} - 1 \right) \label{cond_final} \\
&=& k_BT\,\tau \left[\frac{1}{2} \left(\,\frac{ 2 + \cos(\alpha)}{1- 
\cos(\alpha)}\right)
    + \frac{3}{M} \left(\frac{4}{5} - \frac{1}{4} \,{\csc^2(\alpha/2)} \right) 
+ O\left(\frac{1}{M^2}\right)\right]\;.  	
\end{eqnarray}  
Data for the normalized thermal diffusivity, $D_T\tau/a^2T$, as a function 
of $(\lambda/a)^2$ for $\alpha=90^\circ$ and $M=20$ is compared with 
(\ref{cond_final}) in Fig. \ref{alpha90}b. The agreement is excellent.    
Fig. \ref{cond_corr}a contains a plot of the various contributions to the 
time dependent correlation function $2\langle I_5(0)I_5(t)\rangle/5N(k_BT)^3$, 
and Fig. \ref{cond_corr}b shows the corresponding data for the 
time dependent thermal diffusivity for  
$\alpha=30^\circ$ (filled symbols) and $\alpha=150^\circ$ (unfilled symbols). 
Note that for large collision angles, stress correlations decay very rapidly, 
so that only the first couple of terms in the time series are needed. Finally,  
the normalized thermal diffusivity, $D_T/\tau k_BT$, is plotted as a function 
of the collision angle in Fig. \ref{condlarge} for $M=5$ and $M=20$. 
Again, the results are in excellent agreement with theory. 
It should be emphasized that only the diagonal contributions to $D_T$ have 
been considered here. Although off-diagonal contributions to the thermal 
diffusivity are generally small, better agreement can be achieved 
for $M \leq 10$ if they are included. In particular, these off-diagonal 
contributions are $O(1/M^2)$. They have been calculated explicitly in 
two dimensions in Ref. \cite{ihle_02b}, and for Model B in Sec. \ref{sec:Btd} 
of this paper. 

\subsubsection{Self-Diffusion Coefficient} 

\begin{figure}
\begin{center}
\leavevmode
\includegraphics[width=5in,angle=270]{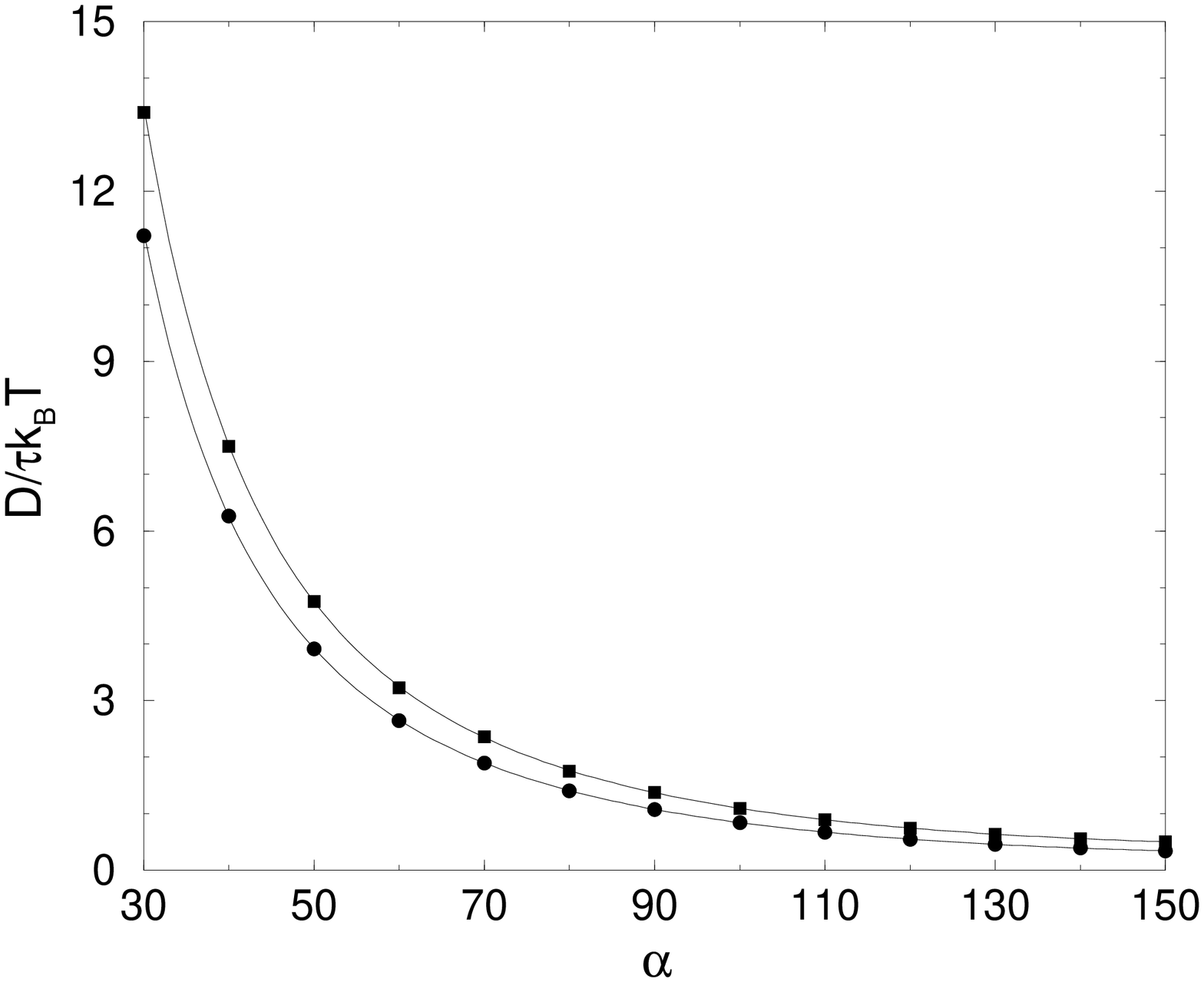}
\caption{Normalized self-diffusion constant, $D/\tau k_BT$, for Model A 
as a function of collision angle $\alpha$. The lines are the theoretical 
prediction, Eq. (\ref{selfdiffusion}). The data were obtained by tme averaging 
over 75000 iterations. Parameters: $L/a=32$, $\lambda/a = 2.309$, 
$\tau=1$, and $M=5$ ($\sblacksquare$) and $M=20$ ($\lbullet$). 
}
\label{diffusion}
\end{center}
\end{figure}

The self-diffusion constant $D$ of particle $i$ is defined by 
\begin{equation} 
D= \lim_{t\to\infty} {1\over 2dt} \langle 
[{\bf r}_i(t)-{\bf r}_i(0)]^2\rangle.   
\end{equation}  
The position of the particle at time $t=n\tau$ is 
\begin{equation} 
{\bf r}_i(t) = {\bf r}_i(0) + \tau \sum_{i=0}^{n-1} {\bf v}_i(k\tau) ,   
\end{equation}  
so that 
\begin{equation} 
\langle [{\bf r}_i(t)-{\bf r}_i(0)]^2\rangle = \tau^2\sum_{j=0}^{n-1}\sum_{k=0}^{n-1}
\langle {\bf v}_i(j\tau)\cdot{\bf v}_i(k\tau) \rangle.    
\end{equation}  
The sums can be rewritten as 
{\setlength\arraycolsep{0pt}
\begin{eqnarray} 
\sum_{j=0}^{n-1}\sum_{k=0}^{n-1}
& &\langle {\bf v}_i(j\tau)\cdot{\bf v}_i(k\tau) \rangle =  
\sum_{j=0}^{n-1} \langle {\bf v}_i^2(j\tau) \rangle + 2 \sum_{j=0}^{n-2} 
\sum_{k=j+1}^{n-1} \langle {\bf v}_i(j\tau)\cdot{\bf v}_i(k\tau) \rangle \nonumber\\     
&=&\ nd\,k_BT + 2\sum_{j=1}^{n-1} j \langle {\bf v}_i(0)\cdot 
{\bf v}_i(n-j)\tau)\rangle. \label{diffsum}       
\end{eqnarray}}    
Expression (\ref{diffsum}) can be evaluated using the same 
approximations as were used to determine the viscosity and thermal 
diffusivity. Setting $d=3$ and using Eq.(\ref{evol}), one gets
\begin{equation}\label{D0tau}  
\langle {\bf v}_i(0)\cdot {\bf v}_i(k\tau)\rangle = 3 k_BT \, \gamma^k\ ,  
\end{equation}   
where 
\begin{equation}\label{gam}
\gamma=[2\cos(\alpha)+1]/3 - 2[\cos(\alpha)-1]/(3M) \;. \label{zetadiff}
\end{equation} 
Substituting Eq.(\ref{D0tau}) into Eq.(\ref{diffsum}), one gets
\begin{eqnarray}
D&=&\lim_{n\to\infty} {k_B T \tau}\left[\frac{1}{2} + \frac{1}{n}
\sum_{j=1}^{n-1}j\gamma^{n-j}\right] \nonumber\\     
&=&\frac{k_BT\tau}{2}\left[\frac{1+\gamma}{1-\gamma}\right],      
\end{eqnarray}
or, explicitly, as a function of $M$,  
\begin{equation}
D=\frac{k_BT\tau}{2}\left[\frac{3}{1-\cos(\alpha)} 
\bigg(\frac{M}{M-1}\bigg) -1  \right].   \label{selfdiffusion}
\end{equation} 
The diffusion coefficient was measured for $M=5$ and $20$ and  
$\lambda/a = 2.309$; the results are shown in Fig. \ref{diffusion}. 

\subsection{Shear viscosity at small mean free path approximation}

Simple kinetic arguments can be used to calculate the rotational 
contribution to the kinematic viscosity \cite{ihle_02a}. 
Consider a collision cell of linear dimension $a$ and divide the cell 
by the plane $z=h$. Since the particle collisions occur in a shifted 
cell coordinate system, they result in a transfer of momentum between 
neighboring cells in the unshifted reference frame. The plane $z=h$ 
represents a cell boundary in the unshifted frame. Consider now the 
momentum transfer in the $z$-direction, and assume 
a homogeneous distribution of particles in the cell. The mean 
velocities in the lower and upper partitions are 
\begin{equation}
{\bf u}_1=\frac{1}{M_1}\,\sum_{i=1}^{M_1}\,{\bf v}_i 
\end{equation}
and 
\begin{equation}
{\bf u}_2=\frac{1}{M_2}\,\sum_{i=M_1+1}^M\,{\bf v}_i\,,  
\end{equation}
respectively, where $M_1=M(a-h)/a$ and $M_2=Mh/a$. 
The $x$-component of the momentum transfer resulting from the collision is 
\begin{equation} 
\Delta p_x(h)\equiv \sum_{i=1}^{M_1}\,[v_{i,x}(t+\tau)-v_{i,x}(t)] \;\;. 
\end{equation} 
Using Eq. (\ref{evolvz}) and averaging over the orientation of the vector 
${\bf{\hat{R}}}$ then yields  
\begin{equation}
\Delta p_x(h)={2 \over 3} (c-1) M_1 (u_{1,x}-u_x) \;\;.
\end{equation}
Since $M{\bf u} = M_1{\bf u}_1 + M_2{\bf u}_2$,  
\begin{equation}
\Delta p_x(h)={2 \over 3} (1-c)\,M\,(u_{2,x}-u_{1,x})\,\,
\frac{h}{a}\,\left(1-\frac{h}{a}\right) \;\;, 
\end{equation}
so that and averaging over $h$---which corresponds to averaging over the 
random grid shift---one has 
\begin{equation}\label{adpx} 
\langle\Delta p_x\rangle= \frac{1}{9}(1-c)M(u_{2,x}-u_{1,x}). 
\end{equation}
Since the dynamic viscosity $\eta$ is defined as the ratio of the tangential 
stress, $P_{zx}$, to $\partial u_x/\partial z$, we have 
\begin{equation}\label{etaap}
\eta = \frac{\langle\Delta p_x\rangle/(a^2\tau)} {\partial u_x/\partial z} = 
       \frac{\langle\Delta p_x\rangle/(a^2\tau)} {(u_{2,x}-u_{1,x})/(a/2)},  
\end{equation}
so that the kinematic viscosity, $\nu=\eta/\rho$, is  
\begin{equation}
\label{rotviscosity}
\nu=\frac{a^2}{18 \tau}[1-\cos(\alpha)]   
\end{equation}
in the limit of small mean free path.

\begin{figure}
\begin{center}
\leavevmode
\includegraphics[width=5in,angle=270]{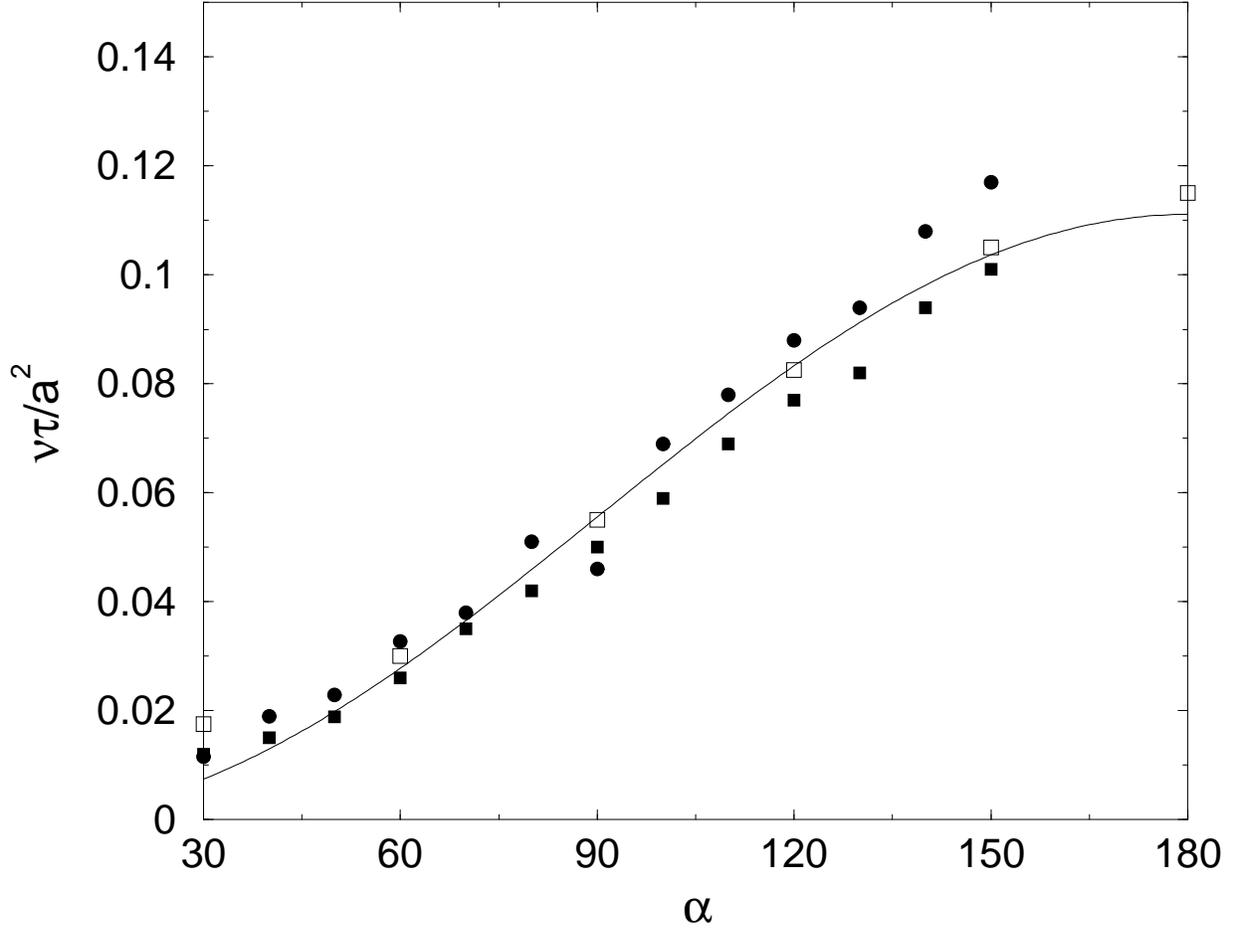}
\caption{Normalized kinematic viscosity for Model A as a function of collision 
angle $\alpha$ for small mean free path, $\lambda/a = 0.0361$.  
The plot shows both the rotational ($\lbullet$) and the total ($\sblacksquare$) contributions to the viscosity. The solid line is the theoretical prediction, 
Eq. (\ref{kin_vis_final}). 
The data were obtained by time averaging over 75000 iterations.
Parameters: $L/a=32$ and $M=20$. The open squares ($\ssquare$) are data for the 
total kinematic viscosity obtained in Ref. \cite{alla_02}. }\label{viscosmall}
\end{center}
\end{figure}

We have measured 
both the rotational and total contributions to the kinematic viscosity 
for $\lambda/a=0.0361$. The results are shown in Fig. \ref{viscosmall}. 
As can be seen, multi-particle collisions provide the dominant contribution 
to the viscosity for small mean free path. Furthermore,  
Eq. (\ref{rotviscosity}) provides an accurate approximation for the viscosity 
in this regime. The systematic deviations for small $\alpha$ are due to 
kinetic contributions (see Fig. \ref{viscolarge}). 

\section{Model B: Rotation around orthogonal axes} \label{sec:reduced}

The second collision rule we will consider involves rotations about one 
of three orthogonal axes. In the implementation considered here, we take 
these axes to be the $x$-, $y$-, and $z$-axes of a cartesian coordinate 
system. At each collision step, one of these axes is chosen at random, and 
a rotation by an angle $\pm\alpha$ is then performed about this axis. 
The sign of $\alpha$ is chosen with equal probability. Rotations 
about the $x$-, $y$-, and $z$-axes are described by the rotation matrices  
\begin{equation}
\label{B_3}
M_x=\left( 
\begin{array}{ccc}
1  & 0  & 0  \\
0  & c  & s  \\
0  & -s & c
\end{array}
\right),\;\;\;
M_y=\left( 
\begin{array}{ccc} 
c  & 0  & s  \\
0  & 1  & 0  \\
-s & 0  & c    
\end{array}
\right),\;\;\;
M_z=\left(
\begin{array}{ccc}
c  & s  & 0  \\
-s & c  & 0  \\
0  & 0  & 1
\end{array}
\right), 
\end{equation}
where $c={\rm cos}(\alpha)$ and $s=\pm{\rm sin}(\alpha)$, depending on the 
sign of $\alpha$. In the following, we will refer to this collision rule 
as Model B. The rate of approach to thermal equilibrium for this model is 
almost identical to that of Model A. This can be seen in Fig. \ref{PRE_3DW_5}, 
which shows the angular dependence of $\tau_R/M$ for two values of 
$\lambda/a$, 1.15 ($\lbullet$) and 0.0361 ($\ssquare$). As in two dimensions, 
the relaxation rate is essentially independent of temperature.   

An advantage of Model B is that the analytical calculations are 
comparatively simple and resemble those for the model in two dimensions.
However, as will be shown in the following section, there are new  
finite cell size corrections which are unique to this collision rule. As 
will be shown, they occur because rotations are performed about one of 
the symmetry axes of the cell lattice.  

\subsection{Large mean free path approximation}

\subsubsection{Shear viscosity}\label{sec:lmfpb}

For large mean free path, we proceed as in Sec. \ref{sec:Asv}. In order to 
determine the shear viscosity in this regime, we need to evaluate 
temporal correlation functions of the type
\begin{equation}\label{B_1}  
A_n=\sum_{ij}\,\langle v_{ix}(0)x_{iy}(0)v_{jx}(n\tau)v_{jy}(n\tau) \rangle. 
\end{equation}
$A_0$ has the same value as in Model A. For $n\ne0$, there are again both 
diagonal ($j=i$) and off-diagonal ($j\ne i$) contributions to $A_n$.  
Using the definition of the rotation 
matrices, Eq. (\ref{B_3}), it is easy to show the diagonal contributions 
to $A_1$ are  
\begin{equation}
\label{B_4}
A_1^x=N(k_B T)^2\zeta_1, 
\end{equation}
\begin{equation}\label{B_5}
A_1^y=N(k_B T)^2\zeta_1,  
\end{equation}   
and 
\begin{equation}\label{B_6}
A_1^z=N(k_B T)^2(\zeta_1^2-\zeta_2^2),
\end{equation}
for rotations around the $x$, $y$ and $z$-axes, respectively, where 
$\zeta_1=1/M+c(1-1/M)$ and $\zeta_2=s(1-1/M)$.
Averaging over the three rotation axes, it follows that the total diagonal 
contribution is 
\begin{equation}
\frac{1}{3}(A_1^x+A_1^y+A_1^z)=N(k_B T)^2\zeta_B ,  
\end{equation} 
where 
\begin{equation} 
\label{B_2}
\zeta_B=(2\zeta_1+\zeta_1^2-\zeta_2^2)/3.  
\end{equation}

The off-diagonal contributions, which come from particles $j$ which are in the 
same cell as particle $i$ at $t=0$, can be evaluated in a similar fashion. The 
result is  
\begin{equation}
\label{B_7}
N(k_BT)^2(M-1)\eta_B,   
\end{equation}
where $\eta_B=2c(c-1)/(3M^2)$, so that 
\begin{equation}
A_1 = N(k_BT)^2[\zeta_B+(M-1)\eta_B].
\end{equation}
The off-diagonal contribution is 
three times smaller than in two dimensions \cite{ihle_02b}.
Note that the leading diagonal contribution is $O(1)$, while that of the 
off-diagonal contribution is $O(1/M)$.

The behavior over longer time intervals can be analyzed in a similar fashion, 
and as for Model A, one finds  
\begin{equation}   
A_n= N(k_BT)^2[\zeta_B+(M-1)\eta_B]^n  
\end{equation}
so that
\begin{equation}\label{mbnu}
\label{kin_vis_B}
\nu = {k_BT\,\tau\over2}\, \left({1+\zeta_B+\eta_B\over1-\zeta_B-\eta_B}\right)
+w_B(c,{a^2\over\tau}) . 
\end{equation}
The last term on the right hand side of Eq. (\ref{kin_vis_B}) is a finite 
cell size correction. In two dimensions and for Model A, $w_A=a^2/(12 \tau)$
\cite{ihle_02b}. As discussed in Sec. \ref{sec:Asv}, it occurs because the 
substitution $\Delta\xi_{iy}=\tau v_{iy}$ in the first term on the sum 
on the right hand side of Eq. (\ref{sv}) is not precisely correct. 
In the present case, however, there are additional corrections because 
the rotation matrices always leave one component of the velocity unaltered.  
As a result, there are contributions to $C_1$ that have projections on $C_0$. 

\bigskip 

\noindent{\em Finite cell size correction:} In order to simplify the 
discussion of the finite cell size corrections, the following calculations 
are performed in the limit $M\rightarrow \infty$. In this case, the time 
evolution equations reduce to  
\begin{eqnarray}
\nonumber        
v_{ix}(\tau)&= & v_{ix}(0) \\
\label{EVOLV11}
v_{iy}(\tau)&= &c v_{iy}(0)+z s v_{iz}(0)
\end{eqnarray}
for rotations around the $x$-axis, 
\begin{eqnarray}
\nonumber
v_{ix}(\tau)&= & c v_{ix}(0)+z s v_{iz}(0) \\
\label{EVOLV12}
v_{iy}(\tau)&= & v_{iy}(0)
\end{eqnarray}
for rotations around the $y$-axis, and  
\begin{eqnarray}
\nonumber
v_{ix}(\tau)&=&c v_{ix}(0)+z s v_{iy}(0) \\
\label{EVOLV13}
v_{iy}(\tau)&=&c v_{iy}(0)-z s v_{ix}(0)
\end{eqnarray}
for rotations around the $z$-axis, where as before, $c$ and $s$ are the 
cosine and the sine of the rotation angle $\alpha$.  
$z=\pm 1$ specifies the sign of $\alpha$. 

When calculating $C_1$, we have to consider rotations about 
the three symmetry axes separately. As can be seen from Eqs. (\ref{EVOLV13}),  
rotations about the $z$-axis mix both the $x$- and $y$-components 
of the velocity, so that the situation is similar to that considered 
in Sec. II B 4 of Ref. \cite{ihle_02b}. Although the same techniques can be 
used to evaluate $C_1^z$ as in Ref. \cite{ihle_02b}, we know from 
the results of that paper that there are no finite cell size corrections in 
this case.  

The situation is different for rotations about the $x$- and $y$-axes. 
For rotations about the $x$-axis, one has  
\begin{equation}
\label{CASEa1}
\tau^2C_1^x=k_B T\sum_i\langle\Delta \xi_{iy}(0) \Delta \xi_{iy}(\tau)\rangle, 
\end{equation}
so that $\langle\Delta \xi_{iy}(0) \Delta \xi_{iy}(\tau)\rangle$ needs to 
be evaluated. Using the approach described in Sec. II B 4  of Ref. 
\cite{ihle_02b}, we have  
\begin{equation}
\label{CASEa2}
\tau^2C^x_1/Nk_BT = a \int_0^a\;dy_0 \sum_{n,m=-\infty}^{\infty}\,
 \int_{(na-y_0)/\tau}^{[(n+1)a-y_0]/\tau}\;dv_y\, \int_{b_0}^{b_1}\; \;dv_z
 nm\, w(v_x) w(v_z)  \,,
\end{equation}
where all velocities are at equal time, so that we have dropped the 
argument $(0)$. Note that the average over $z=\pm 1$ has already be performed.
The limits on the inner integral are
\begin{equation}
b_0 = [(m+n)a-y_0-v_y(1+c)\tau]/(s\tau)
\end{equation}
and
\begin{equation}
b_1 = [(m+n+1)a-y_0-v_y(1+c)\tau]/(s\tau).
\end{equation}
$w(v_x)$ is the Boltzmann distribution,
\begin{equation}
\label{BOLTZ}
w(v_x)=\frac{1}{\sqrt{2 \pi k_B T}}\;{\rm exp}\left\{-\frac{v_x^2}{2k_B
T}\right\}.
\end{equation}
Eq. (\ref{CASEa2}) looks very similar to Eqs. (18) and (41) in \cite{ihle_02b} 
and can be evaluated in an analogous fashion. We therefore only sketch the main 
steps of the analysis, referring to \cite{ihle_02b} for details.

The Poisson sum formula \cite{mors_53}
\begin{equation}
\label{POIS}
\sum_{n=-\infty}^{\infty} g(n)=\sum_{m=-\infty}^{\infty}\int_{-\infty}^{\infty}
g(\phi) {\rm e}^{-2\pi i  m\phi}\,d\phi
\end{equation}
is first used twice to transform the double sum over $m$ and $n$ in Eq. 
(\ref{CASEa2}). Partial integrations over $v_x$ and $v_z$ are then performed.
The temperature independent part of the resulting expression can be determined 
by evaluating the $\bar m = \bar n = 0$ contribution of the remaining sums. 
The final result of these calculations is 
\begin{equation}
\label{FINAL1}
\tau^2C_1^x/ Nk_B T = -{a^2\over12}+O(k_B T).  
\end{equation}

For rotations about the $y$-axis, one has 
\begin{equation}
\label{CASEa3}
\tau^2 C_1^y = k_B T\,c\sum_i  \langle\Delta \xi_{iy}(0) 
\Delta \xi_{iy}(\tau)\rangle. 
\end{equation}
In this case, $v_{iy}(\tau)=v_{iy}(0)$, 
and the calculation of 
$\langle\Delta \xi_{iy}(0) \Delta \xi_{iy}(\tau)\rangle$  
can be performed using the methods outlined above.  
The final result is 
\begin{equation}
\label{FINAL2}
\tau^2C_1^y/N k_B T = -c{a^2\over12}+O(k_B T) . 
\end{equation}
Averaging over all three different rotation axes, it follows that 
\begin{equation}
\label{FINAL3}
\tau^2C_1/N k_B T = -{a^2\over36}(1+c)+O(k_B T). 
\end{equation}
Adding this to the contribution from $C_0$, the final approximation for 
the finite cell size correction for Model B is 
\begin{equation}
\label{B_dis9}
w_B(c,a^2/\tau)={a^2\over18\tau}\left(1-{c\over2}\right) . 
\end{equation}
Although this result is obtained for $M\rightarrow \infty$, and neglects 
contributions from $C_n$ with $n\ge2$, it reproduces the behavior of the 
viscosity over rotation angles between $10^\circ$ and $140^\circ$ and 
$\lambda/a>0.5$ with an error smaller than $2\%$.

For $M\rightarrow \infty$, the off-diagonal contributions to the viscosity 
vanish, and the kinematic viscosity has a minimum at $\alpha=120^\circ$ for 
$\lambda/a\to\infty$. For this value of $\alpha$,  
\begin{equation}
\label{B_L1}
\nu_{min}^B=\tau k_BT\left[\frac{1}{6}+\frac{5}{72}
\left(\frac{a}{\lambda}\right)^2\right]. 
\end{equation}
This is significantly smaller than the minimum value given in Eq. (\ref{amin})
for Model A, but still larger than the minimum value in two dimensions, 
Eq. (\ref{nu2d}). 

\begin{figure}
\begin{center}
\leavevmode
\includegraphics[width=5in,angle=270]{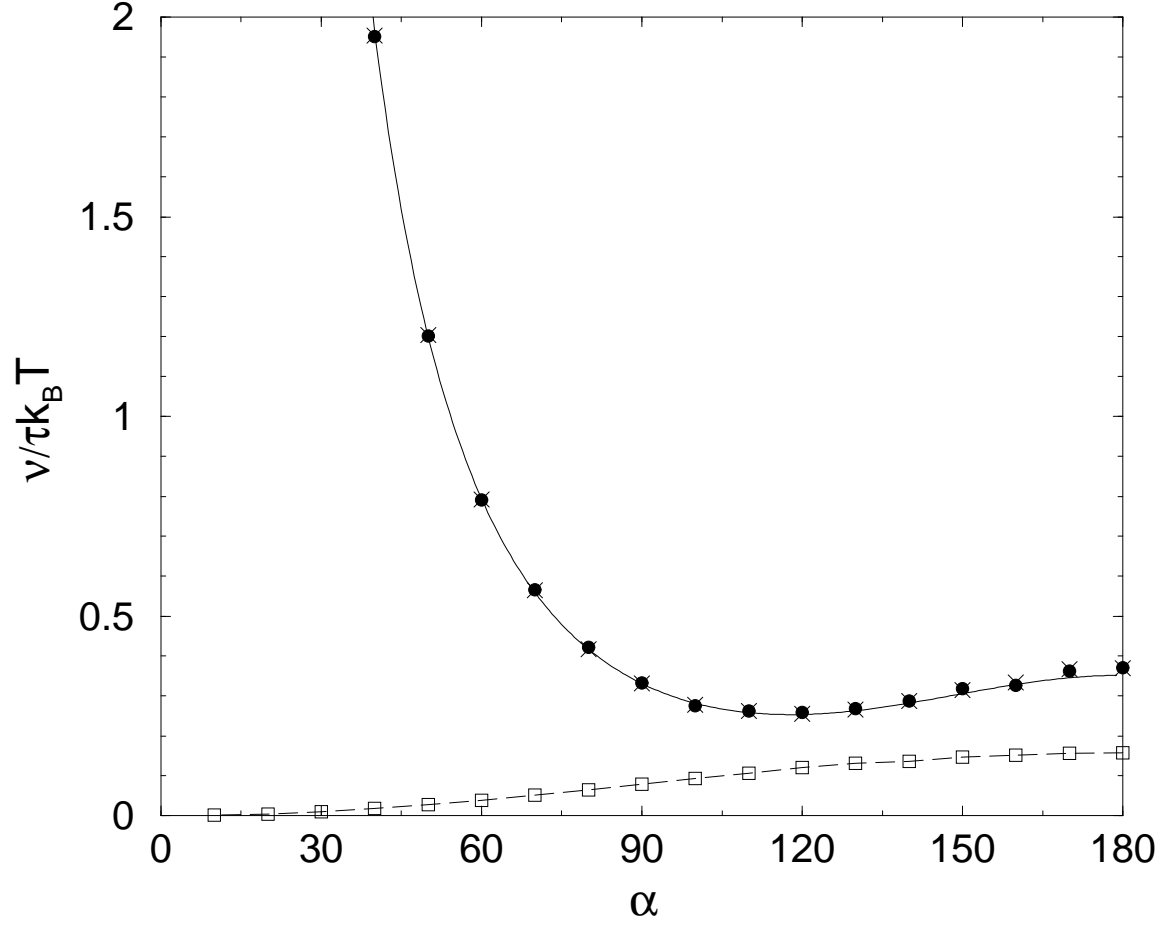}
\caption{Various contributions to the normalized shear viscosity, 
$\nu/\tau k_BT$, for Model B as a function of the rotation angle $\alpha$ 
at large 
mean free path, $\lambda/a=1.15$. The symbols are simulation data for 
the kinetic contribution ($\times$), the rotational contribution 
($\ssquare$), and the total viscosity ($\lbullet$). The solid line is the 
theoretical prediction, Eqs. (\ref{kin_vis_B}) and (\ref{B_dis9}).  
The viscosity has a minimum at $\alpha=120^\circ$, as predicted by theory.
The data were obtained by time averaging over 40000 iterations.
Parameters: $L/a=32$, $\tau=1$, and $M=20$.}\label{PRE_3DW_2a}
\end{center}
\end{figure}

\begin{figure}
\begin{center}
\leavevmode
\includegraphics[width=5in,angle=270]{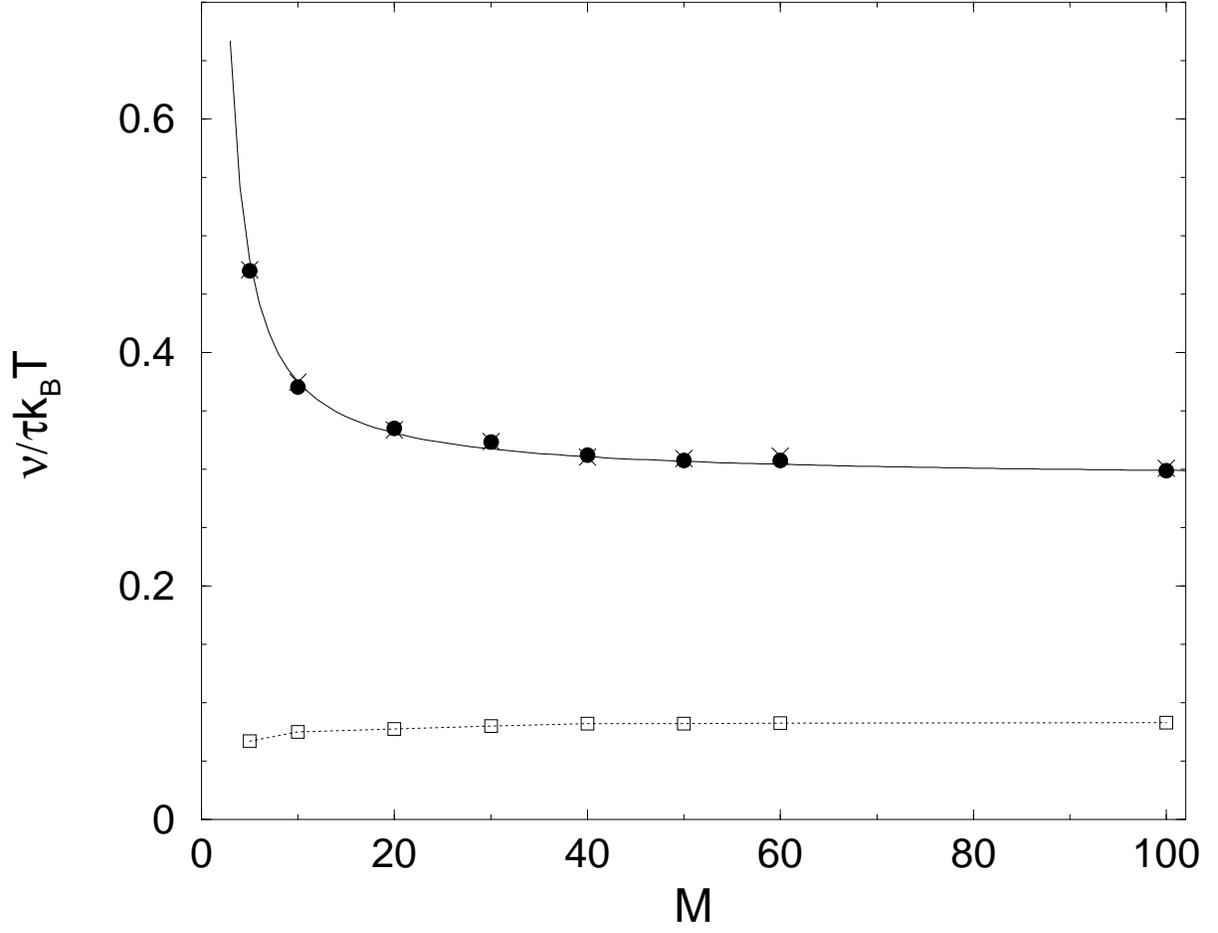}
\caption{Various contributions to the normalized shear viscosity, 
$\nu/\tau k_BT$, for Model B as a function of $M$ for large mean free path, 
$\lambda/a=1.15$, and $\alpha=90^\circ$. The symbols are simulation 
data for the kinetic contribution ($\times$), the rotational contribution
($\ssquare$), and the total viscosity ($\lbullet$). The solid line is the
theoretical prediction, Eqs. (\ref{kin_vis_B}) and (\ref{B_dis9}). 
For this value of $\lambda/a$, rotational contributions to the total viscosity 
are negligible. }\label{PRE_3DW_6}
\end{center}
\end{figure}

\begin{figure}
\begin{center}
\leavevmode
\includegraphics[width=5in,angle=270]{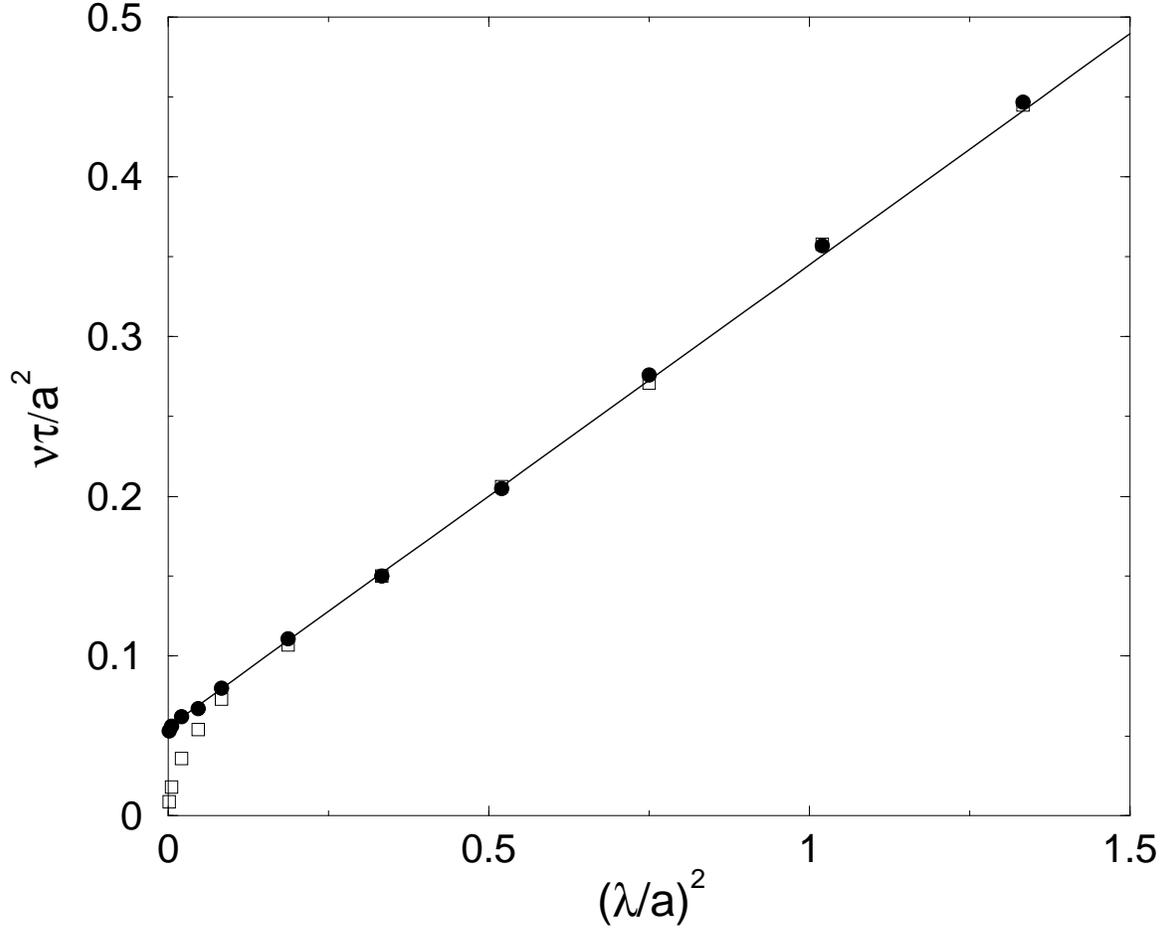}
\caption{
Shear viscosity for Model B as a function of $\lambda/a$ for 
$\alpha=90^\circ$ and  $M=20$. The symbols are simulation data for the 
kinetic contribution ($\ssquare$) and the total viscosity ($\lbullet$). 
The slope of $0.297$ is in excellent agreement with the theoretical 
prediction, 0.2895, which follows from Eqs. (\ref{mbnu}) and (\ref{B_dis9}). 
Note that for $M\rightarrow\infty$, theory predicts a slope of $0.25$; 
$1/M$ corrections are therefore important even for $M=20$.
Parameters: $L/a=32$, $\tau=1$. }\label{PRE_3DW_78}
\end{center}
\end{figure}

Fig. \ref{PRE_3DW_2a} contains a plot of the normalized kinematic 
viscosity, $\nu/(\tau k_BT)$, as a function of $\alpha$ for $\lambda/a=1.15$ 
and $M=20$. 
Data for the the kinetic ($\times$) and rotational ($\ssquare$) contributions, 
as well as the total ($\lbullet$) viscosity, are plotted and compared with the 
theoretical prediction, Eqs. (\ref{kin_vis_B}) and (\ref{B_dis9}). The 
agreement is excellent. Note, in particular, that the finite cell size 
contribution to the total viscosity is not negligible, particularly  
for large rotation angles. In Fig. \ref{PRE_3DW_6}, the normalized viscosity
$\nu/(\tau k_BT)$ is plotted as a function of $M$ for $\alpha=90^\circ$ and 
$\lambda/a=1.15$. Again, the agreement between theory and simulation is 
excellent. Finally, Fig. \ref{PRE_3DW_78} shows the normalized total shear 
viscosity, $\nu\tau/a^2$, as a function of $(\lambda/a)^2$ for $M=20$ 
and $\alpha=90^\circ$. Note 
in particular that the both the $M$ dependence of the viscosity as well 
as the size of the finite cell size correction---given by the intercept---are 
accurately described by theory.   

\subsubsection{Thermal diffusivity}\label{sec:Btd}  

The kinetic part of the reduced flux for the calculation of the thermal 
diffusivity is given by Eq. (\ref{i5}), where again,  
$B_n\equiv\langle I_5^{kin}(\zhat ,0) \vert I_5^{kin}(\zhat ,n\tau)\rangle$.  
The calculation of the thermal diffusivity simplifies considerably if 
we utilize relation (\ref{i6}) and the following relations,      
\begin{equation} 
\left\langle \left(c_pT-{v_i^2(t)\over2}\right) v_{ix}^m \right\rangle 
= 0\ \ \  {\rm for}\ \ \ m=1,2,\ {\rm and}\ \  3,  \\
\end{equation} 
\begin{equation} 
\left\langle \left(c_pT-{v_i^2(t)\over2}\right) v_{ix}^4  
\right\rangle = 3(k_B T)^3, 
\end{equation} 
and 
\begin{equation} 
\left\langle \left(c_pT-{v_i^2(t)\over2}\right)  v_{ix}^2 v_{iy}^2  
\right\rangle = (k_B T)^3. 
\end{equation}  

$B_0$ is the same as in Model A, namely 
\begin{equation}
B_0=\frac{5}{2}N(k_B T)^3. 
\end{equation}
$B_1$ (including off-diagonal terms) is 
\begin{equation}
B_1=\frac{5}{2}N(k_B T)^3\left[\gamma \gamma_4 +(M-1)\gamma_5\right],   
\end{equation} 
where $\gamma$ is given given in Eq. (\ref{gam}),  
\begin{equation}
\gamma_4=[1+2(\zeta_1^2+\zeta_2^2)]/3,   
\end{equation} 
where $\zeta_1$ and $\zeta_2$ are defined in the text following 
Eq. (\ref{B_6}), and 
\begin{equation}
\gamma_5={16\over15}{(1-c)^2\over M^3}.  
\end{equation}

The coefficients $B_n$ form a geometrical series, because successive 
rotations are uncorrelated. This can be seen by {\it first} performing an 
average over the rotation angle and {\it then} performing the thermal 
average. In particular, 
\begin{equation} 
B_n=\frac{5}{2}N(k_B T)^3\left[\gamma \gamma_4 +(M-1)\gamma_5\right]^n,   
\end{equation} 
so that the thermal diffusivity is 
\begin{equation}
D_T={k_BT\tau\over2}\left[ {1+\gamma\gamma_4+(M-1)\gamma_5
\over1-\gamma\gamma_4 -(M-1)\gamma_5}\right] . 
\end{equation}
Note that the off-diagonal contribution is of order $1/M^2$; it is therefore 
less important than for the shear viscosity.  

\subsubsection{Self-Diffusion Coefficient}

The diffusion constant can be determined as in Sec. \ref{sec:tda} for Model
A. The final result is 
\begin{equation}
D={k_BT\tau\over2}\,\left({1+\gamma\over1-\gamma}\right),   
\end{equation}
which is the same as for Model A [see Eq. (\ref{selfdiffusion})]. 
It is interesting to note that 
\begin{equation}
{D_T\over D}\rightarrow 1\;\,{\rm for}\,M\rightarrow\infty, 
\end{equation}
for both Model A and B as well as in two dimensions. 

\subsection{Small mean free path approximation: shear viscosity}

\begin{figure}
\begin{center}
\leavevmode
\includegraphics[width=5in,angle=270]{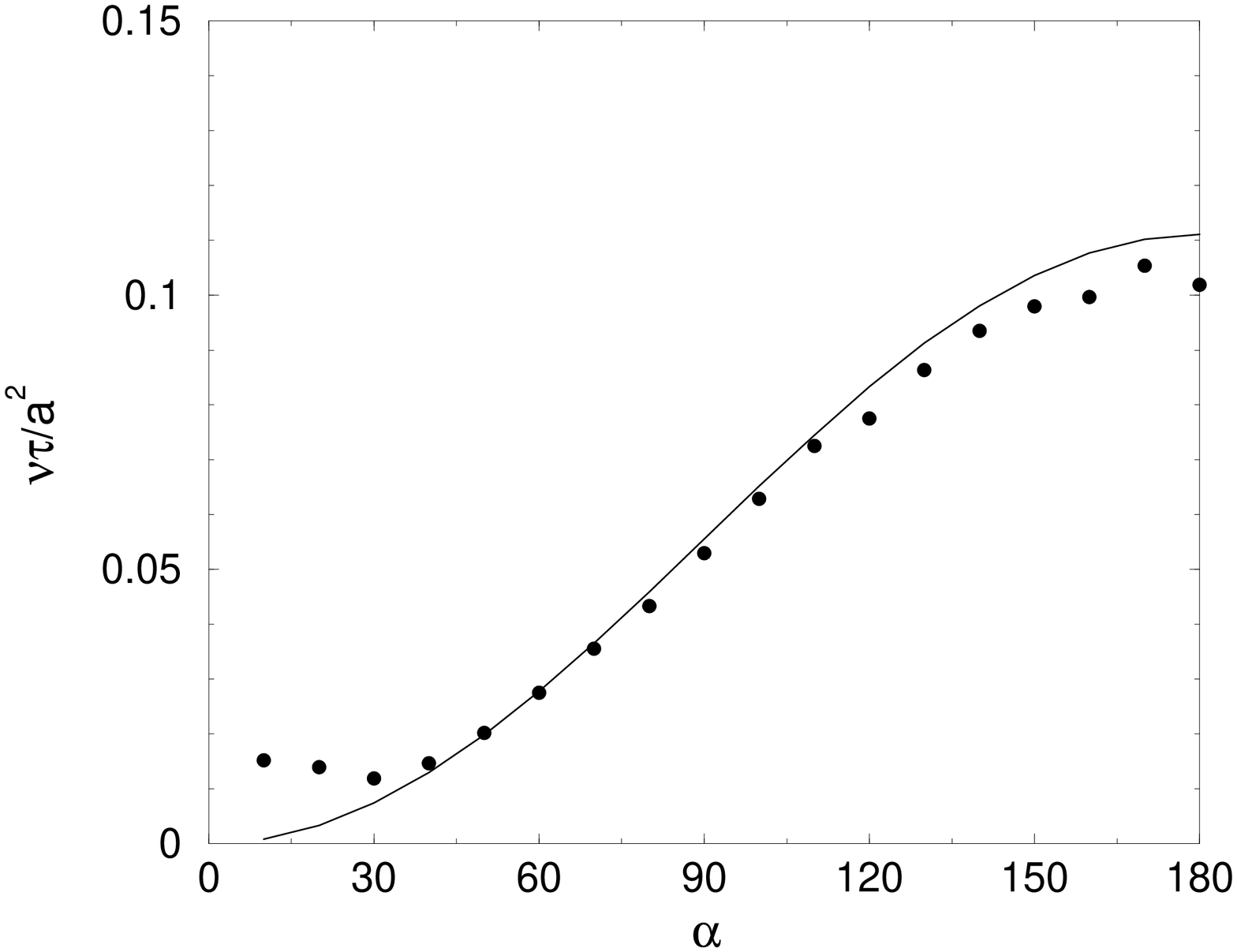}
\caption{Normalized shear viscosity, $\nu\tau/a^2$, for Model B as a function 
of the rotation angle $\alpha$ for small mean free path, 
$\lambda/a=0.0361$. The bullets are simulation data and the solid line is 
the theoretical prediction, Eq. (\ref{B_9}), for the rotational contribution 
to the kinematic viscosity. The deviation of the data from the theoretical 
prediction for $\alpha < 30^\circ$ is due to the increasing importance of the 
kinetic contribution to the viscosity (see Fig. \ref{PRE_3DW_2a}). The data 
were obtained by time averaging over 40000 iterations.  
Parameters: $L/a=32$, $\tau=1$ and $M=20$. }\label{PRE_3DW_1}
\end{center}
\end{figure}

A detailed calculation of the shear viscosity in this limit can be performed 
following the arguments used in Sec. \ref{sec:tda} for Model A and in 
Ref. \cite{ihle_02b} for
two-dimensions. However, for Model B, the following simple argument gives the 
same result. Consider the momentum transfer across a plane perpendicular to 
the $z$-axis. Only rotations about the $x$- and $y$-axes produce a nonzero 
momentum transfer, and since the momentum transfer---and therefore the 
resulting viscosity---from each of these rotations is equal to that calculated 
two-dimensions \cite{ihle_02b}, one finds that  
\begin{equation}
\label{B_9}
\nu_{3D} = \frac{2}{3}\nu_{2D}=\frac{a^2}{18\tau}[1-\cos(\alpha)].  
\end{equation}
Note that this expression is identical to the one obtained for Model A.
Data for the $\alpha$ dependence of the normalized viscosity, $\nu\tau/a^2$, 
at $\lambda/a=0.0361$ is plotted in Fig. \ref{PRE_3DW_1}. Note in particular 
the importance of kinetic contributions to the viscosity for small $\alpha$, 
even for this small value of $\lambda/a$.  

\section{Summary} \label{sec:CONC}

In this paper we have presented a comprehensive analytical and numerical
study of the stochastic rotation dynamics model for fluid dynamics in three
dimensions for two collision rules. The first collision rule (Model A) 
consists of a rotation by an angle $\alpha$ about a randomly chosen axis. 
It was introduced in Refs. \cite{male_99} and \cite{male_00} and used in 
Ref. \cite{alla_02} 
to study channel flow and flow about a spherical object. A new, simpler 
collision rule (Model B), in which collisions involve rotations by an angle 
$\pm\alpha$ about one of three orthogonal axes, was also discussed. 
Calculations involving this model are particularly simple, since the 
rotations about the individual axes are very similar to those in two 
dimensions. In particular, it was possible using this model to calculate 
the off-diagonal contributions to the thermal diffusivity; a similar 
calculation for Model A was prohibitively tedious. Since both models 
are comparable with regard to their computational efficiency, i.e. relaxation 
rates, range of viscosities, etc., the simplicity of Model B can have 
advantages in specific applications.  

Discrete time Green-Kubo relations originally derived in 
Refs. \cite{ihle_01} and \cite{ihle_02a} were 
used to determine explicit expressions for 
the shear viscosity, the thermal diffusivity, and the self-diffusion constant. 
The kinetic, collision, and mixed contributions to the transport coefficients 
were analyzed individually, and no assumptions regarding molecular chaos 
were made. This enabled us to determine correlation induced finite cell 
size corrections to the shear viscosity which persist even in the limit of 
large mean free path. In Ref. \cite{ihle_02b} it was shown that these 
corrections can, under certain circumstances, such as collisions with $\alpha
=90^\circ$ and large particle density, provide the dominant contribution to 
the shear viscosity in two dimensions. In three dimensions, we showed here  
that corrections of this type, while not entirely negligible, are rather 
small for Model A. However, as discussed in Sec. \ref{sec:lmfpb}, for Model B,
where collisions involve rotations about one of three previously defined 
orthogonal axes, there are additional finite cell size corrections which make 
non-negligible contributions to the viscosity for a wide range of densities 
and rotation angles. It is important to note that corrections of this type 
are only important for the shear viscosity. 

It was also shown how quaternion algebra can be used to simplify calculations 
of kinetic contributions to the transport coefficients. In particular, 
the appendicies describe the calculation of the thermal diffusivity in Model A 
using quaternions. Finally, simulation results for 
the viscosity, thermal diffusivity, and the self-diffusion coefficient for 
range of simulation parameters were presented and compared to the analytical 
approximations. In all cases, agreement was excellent; furthermore, the 
comparisons showed that the finite cell size corrections described above 
are necessary in order to achieve quantitative agreement. 

\acknowledgments

Support from the National Science Foundation under Grant No. DMR-0083219, 
the donors of The Petroleum Research Fund, administered by the ACS, the 
Deutsche Forschungsgemeinschaft under Project No. 214283, and 
Sonderforschungsbereich 404 are gratefully acknowledged. We thank E. 
Allahyarov and G. Gompper for providing results of their viscosity 
measurements published in Ref. \cite{alla_02}. 

\appendix
\section{}

The calculation of correlation functions of the reduced fluxes can be 
simplified by rewriting the time evolution equations for the velocities 
using quaternions. Two arbitrary quaternions, $\qP$ and $\qQ$, are defined 
by  
\begin{eqnarray}
\qP &\equiv & (p, {\bf P}) \\
\qQ &\equiv & (q, {\bf Q}), 
\end{eqnarray}
where \{$p$,$q$\} are the scalar parts and \{${\bf P}$,${\bf Q}$\} the 
corresponding vector parts of the quaternions. If the scalar part is zero, 
the quaternion is an ordinary vector and is called as a ``pure'' 
quaternion. The multiplication rule of two quaternions is given by 
\cite{altm_86}
\begin{equation}
\qP \qQ \equiv (pq-{\bf P} . {\bf Q},~ p {\bf Q} + q {\bf P} + 
{\bf P} \times {\bf Q})\;. \label{mrule}
\end{equation} 
It follows that for two pure quaternions, $\qR\equiv(0,{\bf R})$ and 
$\qS\equiv(0,{\bf S})$, 
\begin{equation}
\qR \qS \qR = (0, - |{\bf R} |^2 {\bf S}) \;. \label{mrule2}
\end{equation} 
Defining 
\begin{equation}\label{a1}
\qv (t)  \equiv  (0, \bfv (t)) \;,
\end{equation}
% \noindent
\begin{equation}
\qu \equiv  (0, \bfu_{\bsxi}), 
\end{equation}
and 
\begin{equation}
\qvr \equiv (0, {\bf v^r}) \equiv \qvzero - \qu \;,
\end{equation}
The time evolution equation for the velocities, Eq. (\ref{evol}), can be 
written as  
\begin{equation}
\qvtau = \qa \qvr \qac +\qu \;. \label{qvtau}
\end{equation}                                  
where
\begin{equation}\label{a3}
\qa= (\cos (\alpha/2) , \hat{\bf{R}} \sin (\alpha/2)) \;.
\end{equation}
The first term in Eq. (\ref{qvtau}) corresponds to the rotation of the 
relative velocity vector around the random axis $\hat{\bf{R}}$. Using the 
multiplication rule given in Eq. (\ref{mrule}), it is easy to see that   
Eq. (\ref{qvtau}) is equivalent to Eq. (\ref{evol}). Similarly, 
using Eq. (\ref{mrule2}) it can be shown that 
\begin{equation}
(\qvcubetau)_z = - v^2 (\tau) v_z (\tau) \;. \label{qrel}
\end{equation} 
Dropping the index $i$, $B_1$ given by Eq. (\ref{conduct1}) can be written as 
\begin{equation}
{B_1\over N} = \underbrace{\frac{1}{4} \langle v^2 v^2 (\tau) v_z (\tau) v_z 
\rangle}_{B_1^\prime} -
\underbrace{{c_p T \over 2} \langle v^2 (\tau) v_z (\tau) v_z 
\rangle}_{B_1^{\prime\prime}} , \label{conductapp}
\end{equation}
or by using Eq. (\ref{qrel}), as 
\begin{equation}
{B_1 \over N}= 
{1 \over 4} \left< (\qvcubetau)_z v_z (v_x^2 + v_y^2+v_z^2) \right>
-{c_p T \over 2} \left< (\qvcubetau)_z v_z \right> \;\;.
\end{equation}
Using the multiplication rule for quaternions, and the fact that 
$\qa \qac = 1$, it can be shown that
\begin{eqnarray}
\qvcubetau =&\qa& {\hskip -0.13cm} (\qvr)^3 \qac + \qusq\qa\qvr\qac + 
\qa\qvr\qac\qu\qa\qvr\qac +\qu\qa(\qvr)^2\qac \\
&+& \qa(\qvr)^2\qac\qu+\qucube+\qa\qvr\qac\qusq+\qu\qa\qvr\qac\qu \;\;.
\end{eqnarray}          
Simplifying terms and using energy conservation, 
\begin{equation}
\sum_\alpha [v^r_\alpha(\tau)]^2 - [v^r_\alpha]^2 = 0 \;,
\end{equation}
one obtains 
\begin{eqnarray}
B_1^\prime &=& {1 \over 4} \langle (v_x^2+v_y^2+v_z^2+2 u_{\bsxi x}
(v_x(\tau) -v_x) +2 u_{\bsxi y}(v_y(\tau) -v_y)
+2 u_{\bsxi z}(v_z(\tau) -v_z) )(v_x^2+v_y^2+v_z^2) v_z(\tau) v_z \rangle \\
&=& \frac{(k_BT)^3}{12} \left\{ 35(1+2c) + \frac{2(1-c)}{M}
\left[31-16c+\frac{20}{M}(2c-1) \right] +\frac{576(1-c)^2}{5M^3} \right\}\;\;,
\end{eqnarray}
and
\begin{eqnarray}
B_1^{\prime\prime} &=& {c_p T \over 2} \langle (v_x^2+v_y^2+v_z^2+2 u_{\bsxi x}
(v_x(\tau) -v_x) +2 u_{\bsxi y}(v_y(\tau) -v_y)
+2 u_{\bsxi z}(v_z(\tau) -v_z) )v_z(\tau) v_z \rangle \\
&=& \frac{5(k_BT)^3}{12} \left[5(1+2c)+\frac{10}{M} (1-c) - \frac{16}{5M^2} 
(1-c)^2 (1-{4 \over M}) \right] \;\;,
\end{eqnarray}          
which then yields Eq. (\ref{B1son}) when substituted into 
Eq. (\ref{conductapp}).

% Appendix B

\section{}

In the limit $M \rightarrow \infty$, $\qu\to0$, and Eq. (\ref{qvtau}) can be 
written as 
\begin{equation}\label{a2}
\qvtau = \qa \qv \qac,
\end{equation}
where we have dropped the superscript ``r'', so that 
$\qv \equiv \qv {\hskip -0.05 cm} (0)$. The cube 
of $\qvtau$ is then simply
\begin{equation}\label{a4}
\qvcubetau = \qa \qvcube \qac,
\end{equation}
where
\begin{equation}\label{a5}
\qvcube = (0, - |\bfv |^2 \bfv ) \;.
\end{equation}
This means that $\qvcubetau$ is the rotation of the vector $- |\bfv |^2 \bfv $
around a random axis ${\bf \hat{R}}$.                
Eqs. (\ref{a4}) and (\ref{a5}) can
be used to evaluate the second term in Eq. (\ref{conductapp}), namely 
\begin{equation}
E_1 \equiv \frac{2 B_1^{\prime\prime}}{c_p T}=  \left< v^2(\tau) v_z(\tau) 
v_z \right> = - \left< (\qa \qvcube \qac)_z v_z \right> \;,
\end{equation}
which can be shown to equal
\begin{equation}
E_1= 5 (k_B T)^2 \left[\frac{2 \cos(\alpha)+1}{3}\right] \;\;.
\end{equation}
Similarly, for $t=2\tau$,
\begin{equation}
\qvtwotau = \qap \qvtau \qapc,
\end{equation}
where prime denotes a different random vector then in Eq. (\ref{a3}).
Using energy conservation and the commutator
\begin{equation}
[\qap ,\qv]= (0, 2 \sin(\alpha/2) \; \hat{\bf R^\prime} \times \bfv),  
\end{equation}
$\qvcubetwotau$ can be written as 
\begin{eqnarray}
\qvcubetwotau &=& -|\bfv(2 \tau) |^2  \qvtwotau  \\  
&=& -|\bfv(\tau) |^2  \qap \qvtau \qapc          \\  
&=& -|\bfv(\tau) |^2 (\qv \qap + [\qap , \qv ]) \qapc \\
&=& -|\bfv(\tau) |^2 \qv - |\bfv(\tau) |^2 (0, 2 \sin {\alpha \over 2} \; 
\hat{\bf R^\prime}\times \bfv ) \qapc , 
\end{eqnarray}
so that
\begin{equation}
(\qvcubetwotau )_z = -v^2(\tau) v_z (\tau) 
\left[\frac{2 \cos(\alpha)+1}{3}\right] . 
\end{equation}
Since
\begin{equation}
E_2 \equiv
\left< v^2(2\tau) v_z(2\tau) v_z \right> = - \left< (\qvcubetwotau )_z v_z 
\right>,
\end{equation}
one gets finally, 
\begin{equation}
E_2 = \left[ \frac{2 \cos(\alpha)+1}{3} \right] E_1 = 5 (k_B T)^2 
\left[\frac{2 \cos(\alpha)+1}{3}\right]^2 \;\;.
\end{equation}              
so that the terms $B_n^{\prime\prime}$ form a geometric series. It can also 
be shown that the $B_n^{\prime}$ are terms in a geometric series, with the 
same angular dependence. The difference of these two terms is therefore also 
a geometric series.

\newpage

\newpage

\newpage

\newpage

\newpage

\newpage

\newpage
\bigskip
\newpage

\newpage

\newpage

\newpage

\newpage

\newpage


\begin{thebibliography}{4}       

\bibitem{ihle_01} T. Ihle and D.M. Kroll, Phys. Rev. E {\bf 63}, 020201(R) 
(2001).

\bibitem{ihle_02a} T. Ihle and D.M. Kroll, to appear in Phys. Rev. E.

\bibitem{ihle_02b} T. Ihle and D.M. Kroll, to appear in Phys. Rev. E.

\bibitem{male_99} A. Malevanets and R. Kapral, J. Chem. Phys. {\bf 110}, 
8605 (1999).   

\bibitem{male_00} A. Malevanets and R. Kapral, J. Chem. Phys. {\bf 112}, 
7260 (2000).   

\bibitem{lamu_01} A. Lamura, G. Gompper, T. Ihle, and D.M. Kroll,
Europhys. Lett. {\bf 56}, 319 (2001). 

\bibitem{lamu_02} A. Lamura and G. Gompper, European Phys. J. E {\bf 9}, 
477 (2002).  

\bibitem{alla_02} E. Allahyarov and G. Gompper, Phys. Rev. E {\bf 66}, 
036702 (2002).

\bibitem{male_00a} A. Malevanets and J.M. Yeomans, Europhys. Lett. 
{\bf 52}, 231 (2000).
                         
\bibitem{male_00b} A. Malevanets and J.M. Yeomans, Comput. Phys. 
Commun. {\bf 129}, 282 (2000). 

\bibitem{hash_00} Y. Hashimoto, Y. Chen, and H. Ohashi, Comput. Phys. Comm. 
{\bf 129}, 56 (2000). 

\bibitem{saka_00} T. Sakai, Y. Chen, and H. Ohashi, Comput. Phys. Comm. 
{\bf 129}, 75 (2000). 

\bibitem{saka_02a} T. Sakai, Y. Chen, and H. Ohashi, Phys. Rev. E 
{\bf 65}, 031503
(2002). 

\bibitem{saka_02b} T. Sakai, Y. Chen, and H. Ohashi, Colloids and 
Surfaces A: Physicochem. Eng. Aspects {\bf 201}, 297 (2002).     

\bibitem{inou_01} Y. Inoue, Y. Chen, and H. Ohashi, Comput. Phys. Comm. 
{\bf 142}, 114 (2001). 

\bibitem{inou_02} Y. Inoue, Y. Chen, and H. Ohashi, J. Stat. Phys. 
{\bf 107}, 85 (2002).      

\bibitem{lee_01} S.H. Lee and R. Kapral, Physica A {\bf 298}, 56 (2001). 

\bibitem{mors_53} P.M. Morse and H. Feshbach, {\em Methods of Theoretical
Physics} (McGraw-Hill, New York, 1953).
         
\bibitem{altm_86} S.L. Altman, {\em Rotations, Quaternions, and Double Groups} 
(Clarendon Press, New York, 1986). 

\end{thebibliography}
\end{document}